\documentclass[preprint,eqsecnum]{revtex4}

\usepackage{amsmath,amssymb}
\usepackage{graphicx}
\usepackage{dcolumn}

\newcommand{\nn}{\nonumber}

\begin{document}

\title{\bf{ On the Hamiltonian reduction of geodesic motion on\\
           SU(3)   to SU(3)/SU(2)}}
\author{\sc V.~Gerdt${\,}^a$\,,
            R.~Horan${\,}^b$\,,
            A.~Khvedelidze${\,}^{a,b,c}$\,,
            M.~Lavelle${\,}^b$\,,\\
   \sc      D.~McMullan${\,}^b$\, and
            Yu.~Palii${\,}^a$
\\[0.7cm]
${}^a$ \it Laboratory of Information Technologies, \\
\it Joint Institute for Nuclear Research, \\
\it Dubna, 141980,
\it Russia \\[0.5cm]
${}^b$
\it School of Mathematics and Statistics,\\
\it University of Plymouth,\\
 \it  Plymouth, PL4 8AA, \it United Kingdom\\[0.5cm]
${}^c$
\it Department of Theoretical Physics,\\
\it A.M.Razmadze Mathematical Institute, \\
\it Tbilisi, GE-0193, Georgia}


\begin{abstract}
The reduced Hamiltonian system on
$\mathrm{T}^\ast(\mathrm{SU(3)/SU(2)})$ is derived from a
Riemannian geodesic motion on the SU(3) group manifold
parameterised by the generalised Euler angles and endowed with a
bi-invariant metric. Our calculations show that the metric defined
by the derived reduced Hamiltonian flow on the orbit space
$\mathrm{SU(3)/SU(2)\simeq\mathbb{S}^5}$ is not isometric or even
geodesically equivalent  to the  standard Riemannian metric on the
five-sphere  $\mathbb{S}^5$ embedded into $\mathbb{R}^6$.
\end{abstract}
\maketitle
\newpage

\tableofcontents

\newpage

\section{Introduction}
Symmetry plays a central role in our pursuit of a better
understanding of nature. Through the preservation  or artful
breaking of symmetry, powerful models have been developed which
describe the fundamental forces and which have, so far, withstood
all tests. Indeed, any endeavour to go beyond this standard model
also has, at its heart, an appropriate symmetry argument.

An immediate consequence of symmetry is that it permits for a
reduction in the relevant degrees of freedom needed to describe a
given problem. In a gauge theory this reduction implies that not
all the degrees of freedom present in the formulation of the
theory correspond to physical degrees of freedom. So, for example,
in Quantum Electrodynamics, with its U(1) gauge symmetry, the
potential $A_\mu$, which naively has four degrees of freedom,
describes the photon, which has just two physical degrees of
freedom. Understanding how this type of reduction should best take
place and how the process of quantising a system interacts with
the symmetry, has driven many of the important advances in our
understanding of gauge theories~\cite{'tHooft:2005cq}.

In many cases,  the reduction to the true physical degrees of
freedom in a field theory has been fruitfully  studied through
simpler, finite dimensional systems. In particular, coset spaces of
the form $G/H$, where $G$ and $H$ are finite dimensional Lie groups,
have provided much insight~\cite{Isham:1983zr} into how global and
topological properties of these configuration spaces can be encoded
into the quantisation process via generalised notions of reduction
to the true degrees of freedom~\cite{MT}.

In all investigations to date, specific details on dynamical
aspects of the reduction to $G/H$ have been restricted to  groups
for which manageable parameterisations of the group elements
exist. Essentially this has restricted attention to groups
directly related to the rotation group and its covering,  SU(2).
However, recently there has been much progress in finding suitable
parameterisations for the higher dimensional unitary groups
\cite{TilmaSudarshanSU4,TilmaSudarshanSUN,TilmaSudarshanVol,Bertini}
and particularly for the group SU(3)~\cite{ByrdJMP39,ByrdSudar}.
These advances open the door to detailed investigations of
dynamics on spaces such as the five-sphere, $\mathbb{S}^5$, now
viewed as the reduction from SU(3) to SU(3)/SU(2). By exploiting
our concrete description of this reduction we shall see a new
phenomenon for this system: different metric structures emerge
depending on whether the five sphere is viewed as the coset space
or via its natural embedding in six dimensional Euclidean space.
This is, to the best of our knowledge, the first explicit example
of this metric property of reduction.

The plan of the paper is as follows. We will conclude this
introduction with a brief summary of the classical Hamiltonian
reduction procedure. Then, in section~2, we will see how this
procedure is applied to the group SU(2). This section does not
contain any new results, but fixes notation and introduces themes
that will prepare us for the reduction on the configuration space
SU(3) which will be presented in detail in section~3. Then, in
section~4 we will investigate the possible Riemannian structures
that arise on the quotient space $\mathbb{S}^5$ and discuss the
possible  metric and geodesic correspondences.  In an appendix we
will collect together the details of our consistent parametrisation
of SU(3).

\subsection{Hamiltonian reduction}

Consider the special class of Lagrangian systems whose
configuration space is a compact matrix Lie group $G$. This means
that the state of a system  at fixed time $t=0$ is characterised
by an element of the Lie group $g(0)\in G$ and the evolution is
described by the curve $g(t)$ on the group manifold
\cite{AbrahamMarsden,Arnold}. The ``free evolution'' on the
semisimple group $G$ is, by definition, the Riemannian geodesic
motion on the group manifold with respect to the so-called
Cartan-Killing metric \cite{KobayashiNomizu,Helgason}
\[
 \mathrm{d}s^2_{{}_\mathrm{G}} = \kappa\,
 \mathrm{Tr}\left( g^{-1}\mathrm{d}g \otimes g^{-1}\mathrm{d}g\right)\,,
\]
where $\kappa $ is a normalisation factor.
 The geodesics are given by the extremal curves of the
action functional
\begin{equation}\label{eq;L_g}
S[g]= \frac{\kappa}{2}\,\int^T_{0} \mathrm{d t}\,
\mathrm{Tr}\left(g^{-1} \dot g\,
    g^{-1} \dot{g}\right)\,.
\end{equation}
This action is invariant  under the continuous left translation
$$g(t)\, \to \, g(\varepsilon)\,g(t)\,,
\qquad \varepsilon = (\,\varepsilon_1\,,\varepsilon_2\,,\dots\,,
\varepsilon_{\mathrm{dim}\,\textmd{G}})\,,$$ and therefore  the
system possesses the integrals of motion
$\mathcal{I}_1\,,\mathcal{I}_2\,, \dots\,,
\mathcal{I}_{\,\mathrm{dim}\,\textmd{G}}\,.$ The existence of
these integrals of motion allows us to reduce the number of
degrees of freedom of the system using the well-known method of
Hamiltonian reduction \cite{AbrahamMarsden,Arnold}.

For a generic Hamiltonian system defined on $T^\ast M$ with symmetry
associated to the Lie group $G$ action,  the level set of the
corresponding integrals of motion
\begin{equation}\label{eq:InLevel}
    M_c=\mathcal{I}^{-1}(c)\,.
\end{equation}
where  $c$ is a  set of arbitrary  real constants $c =
(c_1\,,\dots\,, c_{\mathrm{dim}\textmd{G}})$,  determines the
reduced Hamiltonian system on the {\em reduced phase space}
$\mathrm{F}_c \subset M_c$. The subset $\mathrm{F}_c$ is described
by the isotropy group, $G_c\,,$ of the integrals level set $M_c$
\[
\mathrm{F}_c=M_c/G_c\,.
\]
Here we are interested in a special case when the manifold $M$ is
itself a group manifold and the symmetry transformation are group
translations. Now the level set $M_c$ is a subset of the trivial
cotangent bundle $T^\ast G$ which can be identified with the product
of the  group $G$ and its algebra, $G\times g$. The level set given
by the integrals $\mathcal{I}_1=c_1\,,\mathcal{I}_2=c_2\,,\dots\,,
\mathcal{I}_N=c_N\,,\ {N\leq\mathrm{dim}\,\textmd{G}}\,,$ defines
the isotropy group $G_c \subset G$ and the so-called {\em orbit
space}
\begin{equation}\label{eq:coset}
    \mathcal{O} = G/G_c\,.
\end{equation}
The relationship between the  orbit space  $\mathcal{O}$ and  the
reduced phase space $\mathrm{F}_c$ can be summarised as follows
(see e.g. \cite{AbrahamMarsden,Arnold}):
\begin{itemize}
\item {\it the reduced phase space $F_c$  is \underline{
symplectic and diffeomorphic} to the cotangent bundle
$\mathrm{T}^\ast \mathcal{O}$;}
\item {\it the dynamics on the reduced degrees of freedom is
Hamiltonian  with a reduced Hamiltonian given by the projection of
the original Hamiltonian function to $F_c$.}
\end{itemize}
These results are the modern generalisations of the classical
theorems proving that the collection of holonomic constraints
defines a configuration manifold $M$ as  a submanifold of
$\mathbb{R}^n$ and that, in the absence of forces, the trajectories
of mechanical system are geodesics of the induced Riemannian metric.

Note that the above results do not claim that the reduced phase
space and the dynamics on the orbit space are isometric. Indeed, we
know that on the reduced phase space we can define, at least
locally, an induced metric that arises from the kinetic energy
energy part of the reduced Hamiltonian
\begin{equation}\label{eq:Hmm1}
    K_{\mathcal{O}} = \frac{1}{2}\,\mathbf{g}_{\mathcal{O}}(\xi_a,\xi_b)p_a\,p_b\,,
\end{equation}
 On the other hand the map $\pi : G \to G/G_c $ induces the metric
\begin{equation}\label{eq:Hmm2}
 \overline{\mathbf{g}}_{\mathcal{O}}= \pi_\ast
 \mathbf{g}_{{}_\mathrm{G}}\,.
\end{equation}

\noindent We now pose a question about the relation between these
two metrics.
\begin{quote}
{\it When  are the metrics $\mathbf{g}_{\mathcal{O}}$ and
$\overline{\mathbf{g}}_{\mathcal{O}}$  \underline{isometrically} or,
more weakly,  \underline{geodesically} equivalent?}
\end{quote}
We do not know the general answer to this question, so in this
present note we will focus our study on  two  examples: geodesic
motion on the SU(2) and SU(3) group manifolds.

We start with a  well-known example of  Hamiltonian reduction
SU(2) $\rightarrow$ SU(2)/U(1)  and show that the reduced space
{\it is indeed in isometrical correspondence with the cotangent
bundle $\mathrm{T}^\ast \mathbb{S}^2$} and the standard induced
metric on the two-sphere $\mathbb{S}^2$. The case of the SU(3)
$\rightarrow$ SU(3)/SU(2) reduction gives an example of the
opposite result: the metric defined by the Hamiltonian flow on on
the orbit space SU(3)/SU(2) {\it is  not isometrically equivalent}
to a standard round metric on the five-sphere $\mathbb{S}^5$.
Furthermore, in this case, the  stronger result is true: the
reduced configuration space and the standard $\mathbb{S}^5$ are
not even {\it geodesically equivalent}.

\section{Geodesic flow on SU(2)}

In this section we discuss the  example of reduction of free
motion on the SU(2) group manifold. We start with a presentation
of the key geometrical structures found on this group which are
necessary for any further dynamical analysis.

\subsection{The Euler angle parametrisation}

The special unitary group SU(2), considered as a  subgroup of the
general matrix group $\mathrm{GL}(2,\mathbb{C})$, is {\em
topologically} the three-sphere $\mathbb{S}^3$  embedded into
$\mathbb{C}^2$.  This correspondence $\mathrm{SU(2)} \approx
\mathbb{S}^3$  follows immediately from the standard
identification of an  arbitrary element $g \in \mathrm{SU(2)}$ as
\begin{equation}\label{eq:su2mat}
g:=   \left(%
\begin{array}{cc}
  z_1 & -\bar{z}_2 \\
  z_2 &  \bar z_1 \\
\end{array}%
\right)\,, \qquad |z_1|^2+|z_2|^2=1\,.
\end{equation}
The three-sphere $\mathbb{S}^3$  is a manifold which requires more
than one chart to cover it and therefore there is no global
parametrisation of the SU(2) group as a 3-dimensional space. The
local description usually adopted is given by the conventional
symmetric {\em Euler representation } \cite{Euler} for a group
element
\begin{eqnarray}\label{eq:EdSU2}
 \ g\, =\, \exp\left(i\frac{\alpha}{2}\,\sigma_{3}\right)\,\,
           \exp\left(i\frac{\beta}{2}\,\sigma_{2}\right)\,\,
           \exp\left(i\frac{\gamma}{2}\,\sigma_{3}\right)\,,\
\end{eqnarray}
with the appropriately chosen range for  the Euler angles
$\alpha\,,\beta\,,\gamma\,.$

In this representation the generators of the one-parameter subgroups
are the standard Pauli matrices $\sigma_1\,,\sigma_2$ and
$\sigma_3\,,$
\begin{equation}\label{pauli-matr}
 \sigma_{1}=\left(\begin{array}{cc}
   0 & 1 \\
   1 & 0 \
 \end{array}\right),\qquad
 \sigma_{2}=\left(\begin{array}{cc}
   0 & -i \\
   i & \ 0  \
 \end{array}\right),\qquad
  \sigma_{3}=\left(\begin{array}{cc}
   1 & 0  \\
   0 & -1  \
\end{array}\right)\,,
\end{equation}
satisfying the $su(2)$ algebra
\begin{equation}\label{eq:su2alg}
    \sigma_a\sigma_b -\sigma_b\sigma_a = 2\, i\,\epsilon_{abc}\,\sigma_c\,.
\end{equation}
Writing the complex numbers in (\ref{eq:su2mat}) as $z_1=x^1+ i
x^2\,$ and $ z_2= x^3 +i x^4\,$ in polar form
\begin{equation}\label{eq:polcompl}
z_1:=e^{\displaystyle i u}\, \cos{\theta}\,,\qquad
z_2:=e^{\displaystyle i v}\,\sin{\theta}\,,
\end{equation}
and comparing (\ref{eq:su2mat}) with the explicit form of the
Euler matrix (\ref{eq:EdSU2})
\begin{equation}\label{eq:explicitSU2}
 g = \left( \begin {array}{cc}  e^{\displaystyle i\frac{\alpha +\gamma}{2}}\cos
\left( \displaystyle \frac{\beta}{2} \right)
 & e^{\displaystyle i\frac{\alpha -\gamma}{2}}\sin
 \left(\displaystyle\frac{\beta}{2}\right) \\
 \noalign{\medskip}
 - e^{\displaystyle -i\frac{\alpha -\gamma}{2}} \sin
 \left(\displaystyle\frac{\beta}{2} \right)
 & e^{\displaystyle -i\frac{\alpha +\gamma}{2}}
 \cos \left( \displaystyle\frac{\beta}{2}  \right)
\end {array} \right)\,,
\end{equation}
we have
\begin{equation}\label{eq:S5param}
 u = \frac{\alpha +\gamma}{2}\,,
\qquad v=\frac{\alpha -\gamma}{2}\,, \qquad
 \theta= \frac{\beta}{2}\,.
\end{equation}
The Euler decomposition (\ref{eq:EdSU2}) corresponds to the
following parametric representation of the three-sphere embedded in
$\mathbb{R}^4$:
\begin{equation}\label{eq:S3inR4}
    \begin{array}{ccc}
      x^1 &=& \cos\left(\displaystyle \frac{\alpha +\gamma}{2} \right)
      \cos\left( \displaystyle\frac{\beta}{2}\right)\,, \qquad
      x^2 =\sin\left(\displaystyle \frac{\alpha +\gamma}{2}
      \right)\cos\left( \displaystyle\frac{\beta}{2}\right)\,,\\
      &&\\
      x^3 &=& -\cos\left( \displaystyle \frac{\alpha -\gamma}{2}\right)
      \sin\left( \displaystyle\frac{\beta}{2}\right)\,, \qquad
      x^4 = \sin\left(\displaystyle \frac{\alpha -\gamma}{2}
      \right)\sin\left( \displaystyle\frac{\beta}{2}\right)\,.
    \end{array}
\end{equation}
To be more precise, though, this is not a valid parametrisation for
the entire three-sphere. In particular, the neighbourhood of the
identity element of the group in this decomposition turns out to be
degenerate. The identity element of SU(2) corresponds to the whole
set: $\,\beta=0\,$ and $\alpha+\gamma\, =\, 0\,.$ In order to
properly cover the whole group manifold it is necessary to consider
an atlas on the SU(2) group and used different parameterisations on
the different charts. Bearing this in mind, we proceed by assuming
that we are working in a chart $(\mathcal{U}, \phi)$ where $\alpha\,
, \beta$ and $\gamma$ serve as good local coordinates on
$\mathbb{S}^3$ and calculate the Maurer-Cartan forms on SU(2).

Using the following normalisation
\begin{eqnarray}\label{eq:su2Mcn}
g^{-1}\mathrm{d} g \,&=&\, \dfrac{i}{2}
\sum_{a\,=\,1}^3\,\sigma_a \,\otimes\,\omega^a_L \,,\\
\mathrm{d} g\,g^{-1} \,&=&\, \dfrac{i}{2}
\sum_{a\,=\,1}^3\,\sigma_a \,\otimes\,\omega^a_R \,
\end{eqnarray}
and performing the  straightforward calculations with the Eulerian
representation (\ref{eq:EdSU2}) we arrive at the well-known
expressions for left-invariant 1-forms
\begin{eqnarray}
\omega^1_{L}  &=&  \cos\gamma \sin\beta\,
\mathrm{d}\alpha - \sin\gamma\,\mathrm{d}\beta\,,\nn \\
\label{eq:lformSU2} \omega^2_L  &=&  \sin\beta \sin\gamma\,
\mathrm{d}\alpha +
\cos\gamma\,\mathrm{d}\beta\,,\\
 \omega^3_L
&=&\cos\beta\, \mathrm{d}\alpha + \mathrm{d}\gamma\,.\nn
\end{eqnarray}
and the corresponding dual vectors, \(\quad
\omega^a_L(\,X^L_b\,)=\delta_b^a\,, \qquad a,b = 1\,,2\,,3\,, \)
\begin{eqnarray}
X_1^L&=&  \displaystyle{\frac{\cos\gamma}{\sin\beta}}\,
 \frac{\partial}{\partial\alpha}
 -\sin\gamma\,\frac{\partial}{\partial\beta}
-\cot\beta\cos\gamma\,\frac{\partial}{\partial\gamma}\,,
\nn\\
\label{eq:lfieldfSU2}
 X_2^L&=&
\displaystyle{\frac{\sin\gamma}{\sin\beta}}\,\frac{\partial}
{\partial\alpha} +  \cos\gamma\,\frac{\partial}{\partial\beta} -
\cot\beta\sin\gamma\,\frac{\partial} {\partial\gamma}\,,
\\
X_3^L &=& \displaystyle{\frac{\partial}{\partial\gamma}}\,.\nn
\end{eqnarray}
The right invariant 1-forms and the corresponding dual vectors,
\(\quad \omega^a_R(\,X^R_b\,)=\delta_b^a\,,\) are
\begin{eqnarray}
\omega^1_{R} &=& \sin\alpha\,
\mathrm{d}\beta-\cos\alpha\sin\beta\,\mathrm{d}\gamma\,,\nn\\
\label{eq:rformSU2} \omega^2_{R} &=& \cos\alpha\,
\mathrm{d}\beta+\sin\alpha\sin\beta\,\mathrm{d}\gamma\,,
\\
\omega^3_{R} &=&\mathrm{d}\alpha + \cos\beta\,\mathrm{d}\gamma\,.
\nn
\end{eqnarray}
\begin{eqnarray}
X^{R}_1 &= &\cos\alpha\cot\beta\,
\displaystyle{\frac{\partial}{\partial\alpha}} +  \sin\alpha\,
\displaystyle{\frac{\partial}{\partial\beta}} -
\displaystyle{\frac{\cos\alpha}{\sin\beta}}\,
\displaystyle{\frac{\partial}{\partial\gamma}}\,,\nn\\
\label{eq:rfieldSU2}
 X^{R}_2 &=& -
\sin\alpha\cot\beta\,\displaystyle{\frac{\partial}{\partial\alpha}}
+  \cos\alpha\,\displaystyle{\frac{\partial}{\partial\beta}} +
\displaystyle{\frac{\sin\alpha}{\sin\beta}}
\displaystyle{\frac{\partial}{\partial\gamma}}\,,\\
X^{R}_3&=& \displaystyle{\frac{\partial}{\partial\alpha}}\,. \nn
\end{eqnarray}
The  vector fields $X^L_a $  and  $X^R_a$  obey the $su(2)\otimes
su(2)$ algebra with respect to the Lie brackets operation
\begin{eqnarray}\label{eq:algvfLb}
   \left[ X_a^L\,, X_b^L \right]  &=& -\epsilon_{abc}\, X_c^L \\
   \left[ X_a^R\,,  X_b^R \right] &=& \epsilon_{abc}\, X_c^R \\
   \left[ X_a^L\,, X_b^R \right]  &=&  0\,.
\end{eqnarray}

Any compact Lie group can be endowed with the bi-invariant
Riemannian metric build  uniquely (up to a normalization factor)
from the Cartan-Killing form over the algebra. It is convenient to
choose the following normalization for the bi-invariant metric on
the SU(2) group:
\begin{equation}\label{eq:biCK} \mathbf{g}_{{}_{\mbox{\scriptsize
SU(2)}}}= - \frac{1}{2}\,\mathrm{Tr} \left( g^{-1}\mathrm{d} g
\otimes g^{-1}\mathrm{d} g\right)\,.
\end{equation}
In terms of this left/right-invariant non-holonomic frame,
(\ref{eq:biCK}) reads
\begin{eqnarray}\label{eg:metric}
    \mathbf{g}_{{}_{\mbox{\scriptsize SU(2)}}}&=& \frac{1}{4}\,
    \left(\omega^1_L\otimes\omega^1_L
              + \omega^2_L\otimes\omega^2_L
              + \omega^3_L\otimes\omega^3_L \right)\,,\\
&=& \frac{1}{4}\,
    \left(\omega^1_R\otimes\omega^1_R
              + \omega^2_R\otimes\omega^2_R
              + \omega^3_R\otimes\omega^3_R \right)\,.
\end{eqnarray}
Substitution of the expressions (\ref{eq:lformSU2}) and
(\ref{eq:rformSU2}) for the Maurer-Cartan forms $\omega_L$ and
$\omega_R $ yields the metric in the coordinate frame
$\mathrm{d}\alpha\,, \mathrm{d}\beta\,, \mathrm{d}\gamma\,$ basis
\begin{equation}\label{eg:metricS3SU(2)}
    \mathbf{g}_{{}_{\scriptsize \mbox{SU(2)}}} =
                    \frac{1}{4}\,\left(
\mathrm{d}\alpha \otimes \mathrm{d}\alpha + \mathrm{d}\beta
\otimes \mathrm{d}\beta + \mathrm{d}\gamma \otimes
\mathrm{d}\gamma + 2 \cos\beta\, \mathrm{d}\alpha \otimes
\mathrm{d}\gamma\right)\,.
\end{equation}
In order to understand  the {\em metrical} characteristics of a
group manifold viewed as an embedded space, it is instructive to
compare this invariant metric with the metric induced from the
ambient 4-dimensional Euclidian space on the {\em unit} three-sphere
(\ref{eq:S3inR4})
\begin{eqnarray}\label{S3metricinR4}
  \mathbf{g}_{\,{}_{\large{\mathbb{S}^3}}}&=&
   \mathrm{d}\bar{z}_1 \otimes \mathrm{d}{z}_1+
   \mathrm{d}\bar{z}_2 \otimes \mathrm{d}{z}_2\,\\
   &=&\frac{1}{4}\left(\mathrm{d}\alpha \otimes \mathrm{d}\alpha +
 \mathrm{d}\beta \otimes \mathrm{d}\beta + \mathrm{d}\gamma \otimes
\mathrm{d}\gamma + 2 \cos\beta \mathrm{d}\alpha \otimes
\mathrm{d}\gamma \right) \,.\nonumber
\end{eqnarray}
Comparing the metrics, (\ref{eg:metricS3SU(2)}) and
(\ref{S3metricinR4}),  we conclude that the bi-invariant metric on
SU(2) is the same as the standard metric on the unit three-sphere
$ \mathbb{S}^3$  and its bi-invariant volume is
\begin{eqnarray}\label{eq;bivolume}
\mbox{Vol(SU(2))}&=&
   \int \,\sqrt{\det
\mathbf{g}_{{}_{\mbox{\scriptsize SU(2)}}}}\
\mathrm{d}\alpha\wedge
\mathrm{d}\beta\wedge \mathrm{d}\gamma\,\nn\\
    &=&\left(\frac{1}{2}\right)^{3}\int_{0}^{2\pi}  \mathrm{d}\alpha\,
       \int_{0}^{4\pi} \mathrm{d}\gamma\,
\int_{0}^{\pi}\,\mathrm{d}\beta\,\sin(\beta)\, =\,2 \pi^2=
\mbox{Vol}(\,\mathbb{S}^3)\,.
\end{eqnarray}

As a Riemannian manifold the SU(2) group endowed with the metric
(\ref{eg:metricS3SU(2)}) is a 3-dimensional space of constant
curvature with the Riemann scalar
$\mathcal{R}_{{}_{\mbox{\scriptsize SU(2)}}}= 6$ and the Ricci
tensor $\mathcal{R}_{ab}$ given by
\begin{equation}\label{eq:Ricci}
\mathcal{R}_{ab}= \frac{\mathcal{R}_{{}_{\mbox{\scriptsize
SU(2)}}}}{3}\,\mathrm{g}_{ab}= 2\,\mathrm{g}_{ab}\,.
\end{equation}
The Gaussian curvature $K$ of an $n$-dimensional manifold and the
Riemann  scalar  are related via
\begin{equation}\label{eqGC}
K = \frac{\mathcal{R}}{n(n-1)}\,,
\end{equation}
therefore $K_{\scriptsize{\mathrm{SU(2)}}}=1\,$ in agreement with
the volume calculation (\ref{eq;bivolume}).

\subsection{Quotient  SU(2)/U(1)}

Here we recall the key ingredients of  the  construction of a
quotient space $G/H$ by considering the transitive action of the
group $G$ on a certain base space $M$. We have the result that
\footnote{For a rigorous statement we refer to {\it Theorem 3.2}
in \cite{Helgason}.}:
\begin{quote} {\it If the  group $G$ acts
transitively on a set $M$ with $H \subset G$ being an isotropy
subgroup leaving a point  $x_0\in M$ fixed
\[
H=\{g\in G \ |\ g\,\cdot\,x_0= x_0\}\,,
\]
then the set $M$ is in one-to-one correspondence with the  left
cosets $gH$ of $G$\,.}
\end{quote}

The explicit form of this map for the SU(2) group is as follows.
We identify the $su(2)$ algebra  with  $\mathbb{R}^3$ by the map,
$x^a\in\mathbb{R}^3 \to \mathbf{X}\in su(2) \,$
\begin{equation}\label{eq:R3su2}
\mathbf{X}= \sum_{a=1}^{3}\,x^a\sigma_a\,.
\end{equation}
Consider now the {\it adjoint action} of SU(2) on an element of
its algebra $\mathbf{X}\in su(2)\,$
\[
\qquad \mathrm{Ad}(g)(\mathbf{X})=g\,\mathbf{X}\,g^{-1}\,.
\]
The base point $x_0=(0\,,0\,, 1)$ (corresponding to the element
$\sigma_3$) has a one-parameter isotropy subgroup
\[
\mathrm{H}= \exp\left( i\,\dfrac{\alpha}{2}\,\sigma_3\right)\,.
\]
The orbit space of $\sigma_3$
\[
 \mathrm{Ad}(g)(\sigma_3)= g\,\sigma_3\,g^{-1}\,,
\]
is the coset SU(2)/U(1). The proper atlas covering the SU(2) group
manifold provides the coset space parametrisation.
When
$\mathrm{SU(2)}\simeq\mathbb{S}^3$ is parameterised in terms of
two complex coordinates $z_1$ and $z_2$ and the two-sphere is
described by a unit vector $\mathbf{n}= (n^1\,, n^2\,, n^3)$, then
the projection $\mathbb{S}^3 \to \mathbb{S}^2$ reads explicitly
\begin{eqnarray}
(z_1,z_2) \to (n^1\,, n^2\,, n^3)&=& \left( 2\,\Re[\bar{z}_1
z_2]\,,\  2\,\Im[\bar{z}_1{z}_2]\,,\  |z_1|^2-|z_2|^2 \right)\,.
\end{eqnarray}
This is the famous Hopf projection map $ \pi: \mathrm{SU(2)}\, \to
\, \mathbb{S}^2$ showing that SU(2) is a fibre bundle over
$\mathbb{S}^2$ with nonintersecting circles
$\mathrm{U(1)}\equiv\mathbb{S}^1$
 as fibres
\[
\mathbb{S}^1\hookrightarrow \mathrm{SU(2)}
\stackrel{\pi}{\rightarrow }\mathbb{S}^2\,.
\]
Using  the Euler decomposition  (\ref{eq:explicitSU2}) the coset
parametrisation reads
\begin{equation}\label{eq:su2/1}
g\,\sigma_3\,g^{-1}=n^a\,\sigma_a\,,
\end{equation}
with the unit 3-vector
\begin{equation}\label{eq:unitn}
    \mathbf{n}=(-\sin\beta\cos\alpha\,,\,\sin\beta\sin\alpha\,,\,
    \cos\beta\,)\,.
\end{equation}

\subsection{Lagrangian in Euler coordinates}

The bi-invariant Lagrangian
\begin{equation}\label{eq:Lagr2}
    \mathrm{L}_{\scriptsize\mathrm{SU(2)}}=-\frac{1}{2}\,
    \mbox{Tr}\,\bigg( g^{-1}(t)\frac{d}{dt} g(t)
    \,g^{-1}(t)\frac{d}{dt} g(t)\bigg)\,,
\end{equation}
in terms of left/right invariant Maurer-Cartan forms
(\ref{eq:su2Mcn}) reads
\begin{eqnarray}\label{eq:LEul}
    \mathrm{L}_{\mathrm{\scriptsize SU(2)}}&=&
    \frac{1}{4}\,\sum_{a=1}^3\,
    i_{\scriptsize{\dot{U}}}\omega^a_L\,i_{\scriptsize{\dot{U}}}\omega^a_L\,\nn\\
     &=&\frac{1}{4}\,\sum_{a=1}^3\,i_{\scriptsize{\dot{U}}}\omega^a_R\,i_{\scriptsize{\dot{U}}}\omega^a_R\,,
\end{eqnarray}
where $i_{\scriptsize{\dot{U}}}$ is the interior contraction of the
vector field $\scriptsize{\dot{U}}=
\dot\alpha\,\frac{\partial}{\partial\alpha} +
\dot\beta\,\frac{\partial}{\partial\beta}
+\dot\gamma\,\frac{\partial}{\partial\gamma}\,$.

Covering the group manifold with an atlas and considering the chart
where the parameters  $\alpha,\beta , \gamma$  in the Euler
decomposition (\ref{eq:EdSU2}) serve as good coordinates, we arrive
at
\begin{equation}\label{eq:Lcoorrd}
    \mathrm{L}_{\mathrm{\scriptsize SU(2)}}=
    \frac{1}{4}\,\left( \dot\alpha^2+ \dot\beta^2+ \dot\gamma^2 +2\cos(\beta)
    \dot\alpha\dot\gamma \right)\,.
\end{equation}
Comparing (\ref{eq:Lcoorrd}) with the expression
(\ref{eg:metricS3SU(2)}) for the bi-invariant metric on SU(2) we
conclude that
\begin{equation}\label{eq:Lagc}
\mathrm{L}_{\mathrm{\scriptsize SU(2)}}=
\mathbf{g}_{_{\scriptsize{\mathrm{SU(2)}}}}(\dot{U},\dot{U})\,.
\end{equation}

\subsection{Hamiltonian dynamics on $\mathrm{T^\ast SU(2)}$}

The Hamiltonian dynamics on the SU(2) group is defined on the
cotangent bundle $\mathrm{T^\ast SU(2)}$ which can be identified
with the trivialisation $\mathrm{T}^\ast \mathrm{SU(2)}\approx
\mathrm{SU(2)}\times su(2)_L$ or with $\mathrm{T}^\ast
\mathrm{SU(2)}\approx \mathrm{SU(2)}\times su(2)_R\,.$

The canonical Hamiltonian describing geodesic motion on SU(2) can
be obtained by a  Legendre transformation of the Lagrangian
function (\ref{eq:LEul}). Introducing the Poincare-Cartan
symplectic one-form
\[
\Theta= p_\alpha\,\mathrm{d}\alpha + p_\beta\,\mathrm{d}\beta
+p_\gamma\,\mathrm{d}\gamma\,,\] with the canonically conjugated
pairs
$$
\{ \alpha\,, p_\alpha \} = 1\,,\qquad \{ \beta\,, p_\beta \} =
1\,,\qquad\{ \gamma\,, p_\gamma \} = 1\,,
$$
the Hamiltonian on $\mathrm{T^\ast SU(2)}$ is defined as
\begin{eqnarray}\label{eq:H}
    \mathrm{H}_{\mathrm{SU(2)}}&=&
    \,\sum_{a=1}^3\,\xi_a^L\,
   \xi_a^L\,,\nn\\
     &=&\,\sum_{a=1}^3\,\xi_a^R\,
     \xi_a^R\,,
\end{eqnarray}
where $\xi_a^L\,$ and $ \xi_a^R\,$  are the values of the  one-form
$\Theta$ on the left/right invariant vector fields $X_a^L\,,
X_a^R\,$ spanning  the algebra $su(2)_{L,R}$
\[
\xi_a^L: =\Theta\left( X_a^L \right)\, \qquad \xi_a^R: =\Theta\left(
X_a^R \right)\,.
\]
The set of functions $\xi_a^L$  and $\xi_a^R$ obey the
$su(2)_L\,\times\,su(2)_R$ relations  with respect to the Poisson
brackets
\begin{eqnarray}\label{eq:algvf}
  \{\xi_a^L\,, \xi_b^L\} &=& -\epsilon_{abc}\, \xi_c^L \\
   \{\xi_a^R\,, \xi_b^R\}&=& \epsilon_{abc}\, \xi_c^R \\
  \{\xi_a^L\,, \xi_b^R\}&=&0\,.
\end{eqnarray}
In the coordinate frame (\ref{eq:Lcoorrd}) the Hamiltonian
(\ref{eq:H}) becomes
\begin{equation}\label{eq:hamcoor}
 \mathrm{H}_{\mathrm{SU(2)}}=
\frac{p^2_\alpha}{\sin^2(\beta)}+p^2_\beta +
\frac{p^2_\gamma}{\sin^2(\beta)}-\frac{2\,\cos(\beta)}{\sin^2(\beta)
}\,p_\alpha\,p_\gamma\, .
\end{equation}
Now noting that the components of the inverse of the bi-invariant
metric (\ref{eg:metricS3SU(2)}) are
\begin{equation}\label{eq:Lm2}
 \mathrm{g}^{-1}_{{}_\mathrm{SU(2)}}=\frac{4}{
\sin^2(\beta)}\,
\left(%
\begin{array}{ccc}
  1 & 0 & -\cos(\beta) \\
  0 &\sin^2(\beta) & 0 \\
  -\cos(\beta) & 0 & 1 \\
\end{array}%
\right)\, ,
\end{equation}
the Hamiltonian can be rewritten as
\begin{equation}\label{eghsu2f}
    \mathrm{H}_{\,{\mathrm{SU(2)}}}=\frac{1}{4}\
    \mathbf{g}^{-1}_{\,{}_{\scriptsize{\mathrm{SU(2)}}}}(\Theta,\Theta)\,.
\end{equation}

\subsection{Hamiltonian reduction to the coset   {SU(2)}/{U(1)}}

The system with Hamiltonian function (\ref{eq:hamcoor}) has an
obvious first integral
\begin{equation}\label{eq:inp}
    p_\alpha= k\,\qquad\{p_\alpha\,,\mathrm{H}_{\mathrm{SU(2)}}\}
    =0\,,
\end{equation}
where $k$ can be an arbitrary constant. The Hamiltonian on the level
set $M_k:=p_\alpha^{-1}(k)$ is, by definition, the projection of
(\ref{eq:hamcoor}) onto this subspace:
\begin{equation}\label{eq:projh}
\mathrm{H}^{(k)}:= \mathrm{H}_{\mathrm{SU(2)}}\ \biggl|_{
p_\alpha= k}= p^2_\beta + \frac{p^2_\gamma}{\sin^2(\beta)}
-k\,\frac{2\,\cos(\beta)}{\sin^2(\beta) }\,p_\gamma+
\frac{k^2}{\sin^2(\beta)} \,.
\end{equation}
The inverse Legendre transformation gives
\begin{equation}\label{eq:lan}
\mathrm{L}_{\mathrm{{SU(2)}/{SU(1)}}}= \frac{1}{4}
\left(\dot\beta^2+\sin^2(\beta)\,\dot\gamma^2\right) +
k\,\cos(\beta)\dot\gamma\,.
\end{equation}
The interpretation of the system so obtained is the following
\cite{MT}: the first two terms correspond to a particle moving on
the two-sphere $\mathbb{S}^2$ endowed with the standard embedding
metric, while the last term describes the particle interaction
with a Dirac monopole whose potential is
\[
A_\phi:=k\,(1-\cos(\beta))\,.
\]

\section{Geodesic flow on SU(3) using generalised Euler coordinates}


\subsection{Generalised Euler decomposition of SU(3)}

Now we pass on to the description of the Euler decomposition of
the SU(3) group element. The Euler angle parametrisation of the
3-dimensional rotation
 group has been
generalised for the higher orthogonal SO(n) and special unitary
SU(n) groups \cite{TilmaSudarshanSUN,TilmaSudarshanVol,Bertini},
\cite{Murnaghan}, \cite{Wigner} and \cite{VilenkinKlimyk}. Special
attention has been paid to the study of the SU(3)
\cite{BegRuegg,Holland,Yabu,Weigel} and SU(4)
\cite{TilmaSudarshanSU4} groups.

The starting point for the derivation \footnote{We follow the
method of Robert Hermann \cite{Hermann}, who attributed this
construction to C.C. Moore.} of the Euler angle representation of
the  SU(3) group is the so-called Cartan decomposition which holds
for a real semi-simple Lie algebra $\mathcal{G}$. A decomposition
of the algebra $\mathcal{G}$ into the direct sum of vector spaces
$\mathcal{K}$ and $\mathbf{\mathcal{P}}$
\begin{equation}\label{eq;CartDecAlg}
\mathcal{G}=\mathcal{K}\oplus\mathcal{P}
\end{equation}
is a {\em Cartan decomposition of the algebra $\mathcal{G}$} if
\begin{eqnarray}\label{eq:algebra}
&&[\mathcal{K},\mathcal{K} ]  \subset  \mathcal{K}\,,\\
&&[\mathcal{K},\mathcal{P} ]  \subset \mathcal{P}\,,\\
&&[\mathcal{P},\mathcal{P}]  \subset \mathcal{K}\,.
\end{eqnarray}

The Cartan decomposition for a Lie algebra induces a corresponding
{\em Cartan decomposition of the group ${G}$}
\begin{equation}\label{eq:CartDecGroup}
{G}=KP\,,
\end{equation}
where $K$ is a Lie subgroup of $G$ with Lie algebra $\mathcal{K}$
and $P$ is given by the exponential map $P=\exp (\mathcal{P})\,.$

An explicit realisation of the Cartan decomposition for SU(3) can
be achieved using the standard traceless $3\times 3$ Hermitian
Gell-Mann matrices $\lambda_a\,,(a=1\,,\dots\,, 8 )$ (the explicit
form of the $\lambda$ matrices is given in Appendix
\ref{sec:Append1}). Indeed, from the expressions for the
commutation relations
\begin{equation}\label{eq:su(3)algebra}
    [\lambda_a, \, \lambda_b]=2 i\,\sum_{c=1}^{8}\,\mathrm{f}_{abc}\,\lambda_c\,,
\end{equation}
where the structure constants $\mathrm{f}_{abc}$ are antisymmetric
in all  indices and have  the non-zero values
\begin{equation}
 \begin{array}{l}
    \mathrm{f}_{123}=1, \\
    \mathrm{f}_{147}=\mathrm{f}_{246}=\mathrm{f}_{257}=\mathrm{f}_{345}=
    \mathrm{f}_{516}=\mathrm{f}_{637}=1/2,\\
    \mathrm{f}_{458}=\mathrm{f}_{678}=\sqrt{3}/2,
 \end{array}
\end{equation}
it follows  that the set of matrices $(\lambda_1\,, \lambda_2\,,
\lambda_3\,, \lambda_8\, )$ can be used as the basis for the
vector space $\mathcal{K}$ while the matrices  $(\lambda_4\,,
\lambda_5\,, \lambda_6\,, \lambda_7\, )$ span the Cartan subspace
$\mathcal{P}$. Noting that the set of matrices $(\lambda_1,
\lambda_2\,, \lambda_3\,, \lambda_8\, )$ comprise the generators
$(\lambda_1\,, \lambda_2\,, \lambda_3\,)$ of the SU(2) group, one
can locally represent $K$ as the product of the SU(2) subgroup and
a one-parameter subgroup
\begin{equation}\label{eq:K}
    K = \mathrm{SU(2)}\ e^{i \phi \lambda_{8}}\,.
\end{equation}
The second factor, $P=\exp (\mathcal{P})$ in the Cartan
decomposition (\ref{eq:CartDecGroup}) can be represented as a
product of one-parameter subgroups. Moreover,  based on the
algebra (\ref{eq:su(3)algebra}), it can be represented as a
product of a one-parameter subgroup generated by an element
\footnote{ The freedom of choice in the one-parameter subgroups is
analogous to the ``{\em x}'' or  ``{\em y}'' Euler angle
representation of $SU(2)$ with freedom to choose either $\sigma_1$
or $\sigma_2$.} from $\lambda_4\,,\dots \,, \lambda_7$
``sandwiched'' between two different copies of $K$.
 Fixing this generator to be, say,  $\lambda_4$, we have
\begin{equation}\label{eq:P}
    P= K^\prime \ e^{i \theta^\prime\lambda_4 }\ K^{\prime\prime}\,.
\end{equation}
Now observing that $[\lambda_8\,, \lambda_4]= i\sqrt{3}\lambda_5$,
the product $KP$ can be reduced to
\begin{equation}\label{eq:esu3}
    G= \mathrm{SU(2)}\,  e^{i\theta\lambda_5 }\, \mathrm{SU(2)}^{\prime}\,
    e^{i\phi\lambda_8}\,.
\end{equation}
Therefore, finally choosing the Euler representation for the
elements of two subgroups $U\in \mathrm{SU(2)}$ and $V \in
\mathrm{SU(2)}^\prime$ in terms of two sets of angles $(\alpha,
\beta, \gamma)$ and $(a, b, c)$
\begin{eqnarray}\label{eq:subgSU2}
 \ U(\alpha,\beta,\gamma)\, &=&\, \exp\left(i\,\frac{\alpha}{2}\,\lambda_{3}\right)\,\,
           \exp\left(i\,\frac{\beta}{2}\,\lambda_{2}\right)\,\,
           \exp\left(i\,\frac{\gamma}{2}\,\lambda_{3}\right)\,,\\
\qquad \ V(a,b,c)\,& =&\,
\exp\left(i\,\frac{a}{2}\,\lambda_{3}\right)\,\,
                 \exp\left(i\,\frac{b}{2}\,\lambda_{2}\right)\,\,
                 \exp\left(i\,\frac{c}{2}\,\lambda_{3}\right)\,,
\end{eqnarray}
we arrive at the generalised Euler decomposition of an element of
$g \in SU(3)$
\begin{equation}\label{eq:ESU(3)}
    g = U(\alpha,\beta,\gamma)\, Z(\theta,\phi) \, V(a,b,c),
\end{equation}
with
\begin{equation}\label{eq:Z}
    Z(\theta,\phi) := e^{i\,\theta\,\lambda_5 }\, e^{i\,\phi\,\lambda_8}\,.
\end{equation}
Now it is necessary to fix the range of angles in
(\ref{eq:ESU(3)}). Just as in the  case of the SU(2) group where
the Euler parametrisation was  not a global one, the SU(3) group
manifold cannot  be covered by one chart. However there is a range
of parameters such that the parametrisation covers almost the
whole manifold except the set whose measure in the integral
quantities, e.g. such as the invariant volume,  is zero. The
following ranges for the angles in (\ref{eq:ESU(3)})
\begin{eqnarray}\label{eq:range}
&&0 \leq \alpha, a \leq 2\pi \,,\, \,\,\, \qquad
0 \leq \beta\,,  b\,\leq \pi\,,\qquad \, 0 \leq \gamma\,, c\, \leq 4\pi\,, \\
&&\qquad 0 \leq \theta \leq \frac{\pi}{2}\,, \qquad 0 \leq \phi \leq
\sqrt{3}\pi\,,
\end{eqnarray}
lead to the invariant volume for SU(3)
\begin{equation}\label{eq:Vol(SU(3))}
   \mbox{Vol}(SU(3))= \int_{SU(3)}\ast 1 =\sqrt{3}\,\pi^5\,.
\end{equation}
Below this result will be checked by an explicit calculation of
the volume of the SU(3) manifold considered as the Riemannian
space endowed with the bi-invariant metric
\begin{equation}\label{eq:su3ck}
\mathbf{g}_{{}_{\mbox{\scriptsize SU(3)}}}=-
\frac{1}{2}\,\mathrm{Tr}\,\left( g^{-1}\,\mathrm{d}g\otimes
g^{-1}\,\mathrm{d}g\right)\,.
\end{equation}
In terms of the non-holonomic frame built up from the
left/right-invariant forms
\begin{eqnarray}
g^{-1}\mathrm{d} g \,&=&\, \dfrac{i}{2}
\sum_{A\,=\,1}^8\,\lambda_A \,\otimes\,\omega^A_L \,,\\
\mathrm{d} g\,g^{-1} \,&=&\, \dfrac{i}{2}
\sum_{A\,=\,1}^8\,\lambda_A \,\otimes\,\omega^A_R \,,
\end{eqnarray}
the Cartan-Killing metric (\ref{eq:su3ck}) has the diagonal form
\begin{eqnarray}\label{eg:SU3metric}
  \mathbf{g}_{\,{}_{\mbox{\scriptsize{SU(3)}}}}&=&
  \frac{1}{4}\,\left(\,\omega^1_L \otimes\omega^1_L
                    + \omega^2_L \otimes\omega^2_L +
                    \, \dots \,
                    + \omega^8_L \otimes\omega^8_L \, \right)\, \\
&=& \frac{1}{4}\,\left(\, \omega^1_R \otimes\omega^1_R
                    + \omega^2_R \otimes\omega^2_R + \, \dots \,
+ \omega^8_R \otimes\omega^8_R\,  \right)\,,
\end{eqnarray}
while in the corresponding coordinate frame, with the Eulerian
coordinates ($\alpha\,, \beta\,, \gamma\,, a\,, b\,, c\,, \theta\,,
\phi\,)$, presented in Appendix A.2, it becomes
\begin{eqnarray}\label{eg:SU(3)Eulermetric}
\mathbf{g}_{{}_{\mbox{\scriptsize{ SU(3)}}}}&=&
\frac{1}{4}\,\big(\mathrm{d}\alpha \otimes \mathrm{d}\alpha
+\mathrm{d}\beta \otimes \mathrm{d}\beta
+\mathrm{d}\gamma\otimes\mathrm{d}\gamma
+2\cos\beta\,\mathrm{d}\alpha \otimes \mathrm{d}\gamma\big)\nn\\
&+& \frac{1}{4}\,\big( \mathrm{d}a \otimes \mathrm{d}a
+\mathrm{d}b \otimes\mathrm{d}b +\mathrm{d}c \otimes \mathrm{d}c
+2\cos b\,\mathrm{d}a\otimes\mathrm{d}c \big)
\nn\\
&+&\frac{1}{2}\,\cos\theta
\bigg[\sin(a+\gamma)\big(\sin\beta\,\mathrm{d}\alpha \otimes
\mathrm{d}b +\sin b\,\mathrm{d}\beta\otimes \mathrm{d}c\big)\\
&+& \cos(a+\gamma)\big(\mathrm{d}\beta\otimes \mathrm{d}b-
\sin\beta\sin b\,\mathrm{d}\alpha\otimes\mathrm{d}c\big)\bigg]\nn\\
&-&\frac{\sqrt{3}}{2}\sin^2\theta\,\big(\cos\beta\,\mathrm{d}
\alpha+\mathrm{d}\gamma \big)
\otimes \mathrm{d}\phi\nn\\
&+&\frac{1}{4}\,(1+\cos^2\theta)\,\big(\cos\beta\,
\mathrm{d}\alpha+\mathrm{d}\gamma\big) \otimes \big(\mathrm{d}a
+\cos b\,\mathrm{d}c\big) +\mathrm{d}\theta \otimes
\mathrm{d}\theta +\mathrm{d}\phi\otimes \mathrm{d}\phi\,.\nonumber
\end{eqnarray}
Fixing the range of the Euler angles according to (\ref{eq:range})
and noting  that  the determinant of the  Cartan- Killing metric
(\ref{eg:SU(3)Eulermetric}) is \[ \det
\mathbf{g}_{{}_{\mbox{\scriptsize
SU(3)}}}\,=\left(\frac{1}{2}\right)^{12}\,\sin^{6}(\theta)\cos^2(\theta)\sin^{2}(\beta)\sin^{2}(b)\,,
\]
one can check that  the group invariant volume on SU(3) agrees
with (\ref{eq:Vol(SU(3))})
\begin{eqnarray}\label{eq:volumeSU3Euler}
    \mbox{Vol(SU(3))}&=&
    \int_{\mbox{\scriptsize{ SU(3)}}}\, \sqrt{\,\det
\mathbf{g}_{{}_{\mbox{\scriptsize SU(3)}}}}\
\mathrm{d}\alpha\wedge \mathrm{d}\beta\wedge
\mathrm{d}\gamma\wedge \mathrm{d}\theta\wedge
\mathrm{d}a\wedge \mathrm{d}b\wedge \mathrm{d}c \wedge \mathrm{d}\phi\,\nn\\
    &=&\left(\frac{1}{2}\right)^6\,
       \int_{0}^{2\pi}       \mathrm{d}\alpha\,
       \int_{0}^{4\pi}       \mathrm{d}\gamma\,
       \int_{0}^{2\pi}       \mathrm{d}a\,
       \int_{0}^{4\pi}       \mathrm{d}c\,
       \int_{0}^{\sqrt{3}\pi}\mathrm{d}\phi \\
& \times &
       \int_{0}^{\pi}\,      \mathrm{d}\beta\,\sin(\beta)\,
       \int_{0}^{\pi/2}\,    \mathrm{d}\theta\,\cos(\theta)\sin^3(\theta)\,
       \int_{0}^{\pi}\,      \mathrm{d}b\,\sin(b)= \sqrt{3}\,\pi^5\,.\nn
\end{eqnarray}
This volume is in accordance with the general formula established
by I.G.~Macdonald in \cite{MacDonald} and expresses the volume
element of a compact Lie group in terms of the product of volume
elements of odd-dimensional unit spheres
\begin{eqnarray}\label{eq:volumeSU3Mcd}
    \mbox{Vol(SU(3))}
&=&\frac{\sqrt{3}}{2}\times
       \mbox{Vol}(\mathbb{S}^5)\times
       \mbox{Vol}(\mathbb{S}^3)=\frac{\sqrt{3}}{2}\times\pi^3\times2\pi^2 \,.
\end{eqnarray}
In (\ref{eq:volumeSU3Mcd}) the multiplier $\sqrt{3}/2\,,$ comes
from the volume of the maximal torus in SU(3), interpreted
sometimes as the {\em ``stretching''\,} factor
\cite{Bernard,BoyaVolumes}. This fact explicitly shows that the
SU(3) group is not a trivial product of the two spheres,
$\mathbb{S}^3$ and $\mathbb{S}^5$.

The SU(3) group endowed with the bi-invariant metric
(\ref{eg:SU(3)Eulermetric}) has a constant positive Riemann scalar
curvature
\[
\mathcal{R}_{\mathrm{SU(3)}}=24\,,
\]
and the Ricci tensor obeys the relations \footnote{However, in
contrast to the SU(2) group the basic relation defining a space of
constant curvature
$$
\mathcal{R}_{\mu\nu\sigma\lambda} = \frac{\mathcal{R}}{n(n-1)}\,
\left(\mathrm{g}_{\mu\sigma}\mathrm{g}_{\nu\lambda}-
\mathrm{g}_{\mu\lambda}\mathrm{g}_{\nu\sigma}\right)\,
$$
is not valid for the SU(3) group.}
\begin{equation}\label{eq:pc}
\mathcal{R}_{\mu\nu}=\frac{\mathcal{R}_{\mathrm{SU(3)}}}{8}\,
\mathrm{g}_{\mu\nu}= 3\,\mathrm{g}_{\mu\nu}\,.
\end{equation}

\subsection{Geometry of the left coset SU(3)/SU(2)}

The group SU(3) can be viewed as a {\em principal bundle over the
base $\mathbb{S}^5$ with the structure group $\mathrm{SU(2)}$}
\[
SU(2)\hookrightarrow SU(3) \stackrel{\pi}{\rightarrow
}\mathbb{S}^5\,,
\]
with the canonical projection $\pi$ from the SU(3) onto the left
coset SU(3)/SU(2) $\simeq \mathbb{S}^5\,$. This map can be
realised in the following manner. Consider the general linear
group $\mbox{GL}(3, \mathbb{C})$. An arbitrary element
$\mbox{M}_{3\times3}$ can be written in the block form
\begin{equation}\label{eq:matrix3}
\mbox{\large M}_{3\times3} = \left(%
\begin{array}{cc|c}
&&z_3\\
{\raisebox{1.mm}[5mm][5mm]{$\quad \mbox{\large M}_{2\times2}$}}&&z_2  \\
 \hline
 {y}_1~~~~ {y}_2 && {z}_1   \\
\end{array}%
\right)=
\left(%
\begin{array}{cc|c}
&&{\raisebox{-6.mm}{$\mathbf{a}$}}\\
{\raisebox{4.mm}[2mm][2mm]{\quad  $\mbox{\large M}_{2\times2}$}} & &  \\
 \hline
 {\raisebox{-1.mm}[2mm][2mm]{\quad  ${\mathbf{b}}$}}  && {z}_1  \\
\end{array}
\right)\,,
\end{equation}
for complex $2\times 2$ matrix $\mbox{M}_{2\times 2}\,$ and
$z_1\,, z_2\,, z_3\,, y_1\,, y_2 \in\mathbb{C}\,$. The U(3)
subgroup of the $\mbox{GL}(3, \mathbb{C})$ group is defined by the
two matrix equations
\begin{equation}\label{eq:unitar}
\mbox{M}_{3\times 3}\mbox{M}_{3\times 3}^\dag  = \mathbf{I}_{3\times
3}\,, \qquad \mbox{M}_{3\times 3}^\dag\mbox{M}_{3\times 3} =
\mathbf{I}_{3\times3}\,.
\end{equation}
When  $\mbox{M}_{3\times 3}\,$ is represented in block  form,
(\ref{eq:matrix3}), the conditions (\ref{eq:unitar}) reduce  to
the quadratic equations
\begin{eqnarray}\label{eqSU(3)conal}
|z_1|^2+|z_2|^2+|z_3|^2&=& 1\,,\\
|z_1|^2+|y_1|^2+|y_2|^2&=& 1\,,
\end{eqnarray}
and to the set of $2\times2 $ matrix equations
\begin{eqnarray}\label{eqSU(3)conm}
\mbox{M}_{2\times2}\mbox{M}_{2\times2}^\dag + \mathbf{a}\mathbf{a}^\dag
&=& \mbox{I}_{2\times2}\,\\
\mbox{M}_{2\times2}^\dag \mbox{M}_{2\times2} + \mathbf{b}^\dag
\mathbf{b}
&=& \mbox{I}_{2\times 2}\,\\
{z}_1\, \mathbf{a} + \mbox{M}_{2\times2} \mathbf{a}  &=& \mathbf{0}\,\\
\bar{z}_1\,\mathbf{b} + \mbox{M}_{2\times2}^\dag \mathbf{b}  &=&
\mathbf{0}\,.
\end{eqnarray}
Now let $\mathbb{S}^5$ be the five-sphere characterised by a unit
complex vector $\mathbf{Z}:=(z_1\,,z_2\,,z_3)^T\,$
$$
\mathbf{Z}^\dag\mathbf{Z}= 1\,.
$$
The SU(3) group element $g$ then acts on this through left
translations:
\begin{equation}\label{eq:SU3actionS5}
    \mathbf{Z} \rightarrow \mathbf{Z^\prime}= g\mathbf{Z}\,.
\end{equation}
Let $\mathbf{Z}_0$ be the base point on this five-sphere with
coordinates $\mathbf{Z}_0= (0, 0, 1)^T$ whose isotropy group is
\begin{equation}\label{eq:isotropy}
\mbox{\large H}_{3\times3} = \left(%
\begin{array}{cc|c}
&&{\raisebox{-6.mm}{$\mathbf{0}$}}\\
{\raisebox{4.mm}[2mm][2mm]{\quad  $\mbox{\large SU(2)}$}} & &  \\
 \hline
 {\raisebox{-1.mm}[2mm][2mm]{\quad  $\mathbf{0}$}}  && 1  \\
\end{array}
\right)\,.
\end{equation}
Then the coset SU(3)/SU(2) can be identified with the orbit
\begin{equation}\label{eq:su3/2}
\mathbf{Z}= g\cdot(0\,,0\,,1)^T\,.
\end{equation}
Using the explicit form of the representation (\ref{eq:ESU(3)}),
the subgroup SU(2) is embedded into SU(3) as follows:
\begin{equation}\label{eq123}
SU(2)\to SU(3)\,, \qquad V = \left( \begin {array}{ccc}
{e^{\displaystyle -i\,\frac{a +c}{2}} }\displaystyle\cos \left(
\frac{b}{2} \right)
 & -{e^{\displaystyle -i\,\frac{a-c}{2}}
}\displaystyle\sin\left(\frac{b}{2}\right)&0\\
 \noalign{\medskip}
{e^{\displaystyle i\,\frac{a-c}{2}} }\displaystyle\sin\left(
\frac{b}{2} \right)
 & {e^{\displaystyle i\,\frac{a +c}{2}}
}\displaystyle\cos\left(\frac{b}{2}\right)&0\\
\noalign{\medskip} 0&0&1\end {array} \right)\,.
\end{equation}
So the parametrisation of a group element is
\[
g=U\,Z\,V=W\,V\,,
\]
where  the  factor  $W$ reads
\begin{equation}\label{eq:45678}
W= \left(
\begin {array}{ccc}
\displaystyle\cos\theta\cos\frac{\beta}{2}\, e^{\displaystyle i( u
+\frac{1}{\sqrt{3}}\phi )}&
 \displaystyle\sin\frac{\beta}{2}\
 e^{\displaystyle {i}(v+\frac{1}{\sqrt{3}}\phi)}&
\displaystyle\sin\theta\cos\frac{\beta}{2}\, e^{\displaystyle
{i}(u-\frac{2}{\sqrt{3}}\phi)}
 \\
 \noalign{\medskip}
 \displaystyle -\cos\theta\sin\frac{\beta}{2}\,
 e^{\displaystyle -{i}
 (v-\frac{1}{\sqrt{3}}\phi)}&
 \displaystyle \cos\frac{\beta}{2}\, \displaystyle
 e^{\displaystyle -{i}(u-\frac{1}{\sqrt{3}}
 \phi)} &
 -\displaystyle\sin\theta\sin\frac{\beta}{2}\,
e^{\displaystyle-{i}(v+\frac{2}{\sqrt{3}}\phi)}\\
 \noalign{\medskip}
  \displaystyle -\sin\theta\,e^{\displaystyle
 \frac{i}{\sqrt{3}}\phi}&
 0
 &
\displaystyle \cos\theta\ e^{\displaystyle
 -i\frac{2}{\sqrt{3}}\phi}
\end {array}
 \right)\,.\nonumber
\end{equation}
\[
u=\frac{\alpha+\gamma}{2}\,,\qquad v=\frac{\alpha-\gamma}{2}\,.
\]
Using these representations in (\ref{eq:su3/2}) we easily identify
the projection onto the left coset as a five-sphere:
$$\pi:g\in SU(3) \to (z_1\,,z_2\,,z_3)\in \mathbb{S}^5 \,,  $$
 which explicitly reads
 \begin{eqnarray}
 z_1&=&\ \ \displaystyle  \cos\theta\,
 e^{\displaystyle -i\frac{2}{\sqrt{3}}\,\phi}\,,\\
 z_2&=&-\displaystyle \sin\theta\sin\frac{\beta}{2}\
  e^{\displaystyle-\frac{i}{2}(\alpha-\gamma+\frac{4}{\sqrt{3}}\,\phi)}
 \,,\\
z_3&=&\displaystyle \sin\theta\cos\frac{\beta}{2}\
 e^{\displaystyle \frac{i}{2}(\alpha + \gamma-\frac{4}{\sqrt{3}}\,\phi)}\,.
 \end{eqnarray}
Under this projection the Euclidean metric
$\,\mbox{tr}(\mathrm{d}M\mathrm{d}M^\dag)\,$ on GL(3,$\mathbb{C}$)
induces  the following metric on  a unit $\mathbb{S}^5$
\begin{eqnarray}\label{eq:S5eu}
 \mathbf{g}_{\,{}_{\scriptsize{\mathbb{S}^5}}}&=&
 \mathrm{d}\bar{z}_1\otimes \mathrm{d}{z}_1+
\mathrm{d}\bar{z}_2 \otimes\mathrm{d}{z}_2+
\mathrm{d}\bar{z}_3\otimes
\mathrm{d}{z}_3\\
   \nn\\
  &=& \sin^2\theta\bigg(\frac{1}{4}\,\big(
  \mathrm{d}\alpha\otimes\mathrm{d}\alpha+
  \mathrm{d}\beta\otimes\mathrm{d}\beta +
  \mathrm{d}\gamma\otimes\mathrm{d}\gamma+
  2\cos\beta\,\mathrm{d}\alpha\otimes\mathrm{d}\gamma\big)\nonumber \\
  &-&\frac{2}{\sqrt{3}}\,(\cos\beta\mathrm{d}\alpha+\mathrm{d}\gamma)
  \otimes\mathrm{d}\phi\,\bigg)
+\mathrm{d}\theta\otimes\mathrm{d}\theta +
  \frac{4}{3}\,\mathrm{d}\phi\otimes\mathrm{d}\phi\,,\nn
\end{eqnarray}
whose determinant is
\begin{equation}\label{eq:detS5}
\det \mathbf{g}_{{}_{\scriptsize{\mathbb{S}^5}}}\,= \frac{1}{48}\,
\sin^{6}(\theta)\cos^2(\theta)\sin^{2}(\beta)\,.
\end{equation}

The metric (\ref{eq:S5eu}) defines a unit five-sphere
$\mathbb{S}^5$ as a constant curvature Riemann manifold
\begin{equation}\label{eq:S5curvature}
    \mathcal{R}_{\mathbb{S}^5} = 20\,,
\end{equation}
 which is in accordance with its
Gaussian curvature
\[
K_{\mathbb{S}^5}=\frac{\mathcal{R}_{\mathbb{S}^5}}{5(5-1)}=1\,,
\]
as well as with its  volume
\begin{eqnarray}\label{eq:volumeS5}
    \mbox{Vol}(\mathbb{S}^5)&=&
    \int_{\mathbb{S}^5}\, \sqrt{\,\det
\mathbf{g}_{{}_{\scriptsize{\mathbb{S}^5}}}}\
\mathrm{d}\alpha\wedge \mathrm{d}\beta\wedge
\mathrm{d}\gamma\wedge \mathrm{d}\theta\wedge
\mathrm{d}\phi\,\\
    &=&\frac{1}{4\sqrt{3}}\,
       \int_{0}^{2\pi}       \mathrm{d}\alpha\,
       \int_{0}^{4\pi}       \mathrm{d}\gamma\,
       \int_{0}^{\sqrt{3}\pi}\mathrm{d}\phi
       \int_{0}^{\pi}\,      \mathrm{d}\beta\,\sin(\beta)\,
       \int_{0}^{\pi/2}\,    \mathrm{d}\theta\,\cos(\theta)\sin^3(\theta)\,
              = \pi^3\,. \nn
\end{eqnarray}

\subsection{Lagrangian on SU(3) in terms of generalised Euler angles}

Consider the Lagrangian describing the geodesic motion on the  SU(3)
group manifold with respect to the bi-invariant metric
(\ref{eq:su3ck})
\begin{equation}\label{eq:Lagr}
    \mathrm{L}_{\scriptsize\mathrm{SU(3)}}=-\frac{1}{2}\,
    \mbox{Tr}\,\bigg( g^{-1}(t)\frac{d}{dt} g(t)
    \,g^{-1}(t)\frac{d}{dt} g(t)\bigg)\,.
\end{equation}
Using the generalised Euler angles on SU(3) as the configuration
space coordinates
 and (\ref{eg:SU(3)Eulermetric}) for
the bi-invariant metric, one can write the Lagrangian
(\ref{eq:Lagr}) as
\begin{eqnarray}\label{eg:SU(3)LagEulermetric}
\mathrm{L}_{\scriptsize\mathrm{SU(3)}}
&=&\frac{1}{4}\,\big(\dot\alpha^2 +\dot\beta^2+\dot\gamma^2
+2\cos\beta\,\dot\alpha \dot\gamma+ \dot{a}^2 +\dot{b}^2+\dot{c}^2
+2\cos b\,\dot{a}
\dot{c}\big) \\
 &+&\frac{1}{2}\cos\theta\,
\bigg(\sin(a+\gamma)\big(\sin\beta\,\dot\alpha \dot{b} +\sin
b\,\dot\beta\dot{c}\big)+ \cos(a+\gamma)\big(\dot\beta\dot{b}-
\sin\beta\sin b\,\dot\alpha\dot{c}\big)\bigg)\nn\\
&-&\frac{\sqrt{3}}{2}\sin^2\theta\big(\cos\beta\,\dot\alpha+\dot\gamma
\big)\dot\phi +\frac{1}{4}\,(1+\cos^2\theta)\big(\cos\beta\,
\dot\alpha+\dot\gamma\big)\big(\dot{a} +\cos b\,\dot{c}\big)
+\dot{\theta}^2 + \dot{\phi}^2\,.\nn
\end{eqnarray}
From this expression and (\ref{eg:SU(3)Eulermetric}) for it
follows that
\begin{equation}\label{eq:lsu(3)ci}
\mathrm{L}_{\mathrm{\scriptsize SU(3)}}=
\mathbf{g}_{_{\scriptsize{\mathrm{SU(3)}}}}(\dot{Z},\dot{Z})\,.
\end{equation}
where $\dot{Z}$ is the vector field on the tangent bundle TSU(3)
\begin{eqnarray}
{\dot{Z}}= \dot\alpha\,\frac{\partial}{\partial\alpha} +
\dot\beta\,\frac{\partial}{\partial\beta}
+\dot\gamma\,\frac{\partial}{\partial\gamma}
+\dot{\theta}\,\frac{\partial}{\partial \theta} +
\dot{\phi}\,\frac{\partial}{\partial \phi}
+\dot{a}\,\frac{\partial}{\partial a} +
\dot{b}\,\frac{\partial}{\partial b}
+\dot{c}\,\frac{\partial}{\partial c}\,.
\end{eqnarray}

It is worth to note that the Euler decomposition (\ref{eq:ESU(3)})
for elements of SU(3) in terms of the SU(2) subgroups,
\[
\mathrm{SU(3)}=U(\alpha, \beta, \gamma)\,
\exp(i\,\theta\,\lambda_5)\, V(a,b,c)\,
\,\exp(i\,\phi\,\lambda_8)\,,
\]
allows for the expression of the SU(3) Lagrangian
(\ref{eg:SU(3)LagEulermetric}) in terms of the corresponding left
and right invariant elements of the SU(2) Maurer-Cartan 1-forms:
\begin{eqnarray}\label{eq:su3insu2}
\mathrm{L}_{\scriptsize\mathrm{SU(3)}}&=& \frac{1}{4}
\sum_{a=1}^{3}\,i_{\scriptsize{\dot{U}}}\omega^a_L\,i_{\scriptsize{\dot{U}}}
\omega^a_L +
\frac{1}{4}\sum_{a=1}^{3}\,i_{\scriptsize{\dot{V}}}\omega^a_L\,
i_{\scriptsize{\dot{V}}}\omega^a_L\nonumber\bigg.\\
&+&\frac{1}{2}\cos\theta\,\sum_{a=1}^{2}\,
i_{\scriptsize{\dot{U}}}\omega^a_L\,i_{\scriptsize{\dot{V}}}\omega^i_R
-\frac{1}{4}(1+\cos^2\theta)\,
i_{\scriptsize{\dot{U}}}\omega^3_L\,i_{\scriptsize{\dot{V}}}\omega^3_R
\bigg.\nn\\
&-&\frac{\sqrt{3}}{2}\,\sin^2\theta\,
i_{\scriptsize{\dot{U}}}\omega^3_L\, \dot{\phi} +\dot{\theta}^2 +
\dot{\phi}^2\,.
\end{eqnarray}
Here $i_{\scriptsize{\dot{U}}}$ and $i_{\scriptsize{\dot{V}}}$
denote  the interior contraction of the  vector field on each copy
of the SU(2) group,  $U$ and $V$ respectively
\begin{eqnarray}
\scriptsize{\dot{U}}= \dot\alpha\,\frac{\partial}{\partial\alpha}
+ \dot\beta\,\frac{\partial}{\partial\beta}
+\dot\gamma\,\frac{\partial}{\partial\gamma}\,, \qquad
\scriptsize{\dot{V}}= \dot{a}\,\frac{\partial}{\partial a} +
\dot{b}\,\frac{\partial}{\partial b}
+\dot{c}\,\frac{\partial}{\partial c}\,.
\end{eqnarray}

\subsection{Hamiltonian dynamics on SU(3)}

Performing the Legendre transformation, we derive the canonical
Hamiltonian generating the dynamics on the SU(3) group manifold:
\begin{eqnarray}\label{eq:explsu3H}
H_{\scriptsize\mathrm{SU(3)}} &=& \frac{1}{\sin^2{\theta}}
 \left[ \frac{p_\alpha^2}{\sin^2\beta}
 +{p_\beta^2}+
 \left(\tan^2\,{\theta}+\frac{1}{\sin^2\beta}\right)p_\gamma^2
 -
2\,\frac{\cos\beta}{\sin^2\beta}\ p_\alpha p_\gamma \right.
  \\
&+&
 \left.\sin^2{\theta}\left(1+\frac{1}{4}\cot^2\theta+\frac{1}{\sin^2 b}
 \right)p_a^2
 +{p_b^2}+
\frac{1}{\sin^2 b}p_c^2 - 2\,\frac{\cos b}{\sin^2 b}\
p_a p_c \right]\nn\\
&+&\bigg.
   2\,\frac{\cos\theta}{\sin^2\theta\sin\beta\sin b}
\bigg[\cos(a+\gamma)\bigg((p_\alpha -\cos\beta\,p_\gamma)(p_c-
\cos b\, p_a) - \sin b\,p_\beta\,p_b\bigg)\bigg.\nn\\
& -&\bigg.\sin(a+\gamma)\bigg(\sin b(p_\alpha-\cos\beta\,
p_\gamma)p_b+\sin\beta(p_c-\cos b\,p_a)p_\beta\bigg)\bigg]\nn\\
   &+&\frac{1}{4}\,p^2_\theta + \frac{1}{16}\left(
1+\frac{3}{\cos^2\,{\theta}}
   \right)\ p^2_\phi +
\frac{\sqrt{3}}{2}\frac{p_\gamma
p_\phi}{\cos^2\theta}-\frac{\sqrt{3}}{4}
\left(1+\frac{1}{\cos^2\theta}\right) p_a p_\phi\,.\nn
\end{eqnarray}
The Hamiltonian (\ref{eq:explsu3H}) can be rewritten in a compact
form using the left and right-invariant vector fields on the two
SU(2) group copies, $U$ and $V$ used in the Euler decomposition
(\ref{eq:ESU(3)}):
\begin{eqnarray}\label{eq:SU3LR}
  H_{\scriptsize\mathrm{SU(3)}} &=& \sum_{a=1}^3\, \zeta_a^R \,\zeta_a^R+
 \frac{1}{\sin^2\theta}\,\sum_{a=1}^2\,(\xi_a^L-\cos\theta\,\zeta_a^R)^2
\bigg.
 \nonumber \\
  &+&\bigg.\,\frac{1}{\sin^22\,\theta}(2\,\xi_3^L-(1+\cos^2\theta)\,\zeta_3^R-
\frac{\sqrt{3}}{2}\sin^2\theta\,p_\phi)^2+\frac{1}{4}\,p^2_\theta
+\frac{1}{4}\,p^2_\phi\,.
\end{eqnarray}
Here $\xi_a^L$ and $\zeta_a^R$ are functions defined through the
relations
\[
\xi_a^L: =\Theta\left( X_a^L \right)\,, \qquad \zeta_a^R:
=\Theta\left( Y_a^R \right)\,,
\]
with the  SU(2) left-invariant vector fields $X_a^L$ on the tangent
space to the $U$ subgroup, $TU$, and the right-invariant fields
$Y_a^R$ on $TV$ correspondingly.

\subsection{Hamiltonian reduction to SU(3)/SU(2)}

The representation (\ref{eq:SU3LR}) is very convenient for
performing the reduction in degrees of freedom associated with the
SU(2) symmetry transformation. Due to the algebra of Poisson
brackets (\ref{eq:algvf}) the functions $\zeta_1^L\,,\zeta^L_2\,$
and $\zeta_3^L\,$ are the first integrals
\[
\{\zeta^L_a\,, H_{\scriptsize\mathrm{SU(3)}} \}= 0\,.
\]
Let us consider the zero level of these integrals
\begin{eqnarray}\label{eq:SU20}
    \zeta_1^L=0\,,\qquad \zeta^L_2=0\,,\qquad \zeta^L_3=0\,.
\end{eqnarray}
Noting the relation between the left and right invariant vector
fields on a group one can express the functions $\zeta^R_a$
entering in the Hamiltonian as
\[
\zeta^R_c= \mathrm{Ad(V)}_{cb}\,\zeta^L_b\,,
\]
where  $\mathrm{Ad(V)}_{cb}$ is an adjoint  matrix of an element
$\mathrm{V} \in \mathrm{SU(2)}$. From this one can immediately
find the reduced Hamiltonian on the integral level
(\ref{eq:SU20}). Indeed, projecting the expression
(\ref{eq:SU3LR}) on $\zeta_a^R=0$ we find
\begin{eqnarray} \label{eq:SU3/SU2}
 H_{\scriptsize\mathrm{SU(3)/SU(2)}} &=&
  \frac{1}{\sin^2\theta}\sum_{a=1}^3\,\xi_a^L\,\xi_a^L
+
\frac{1}{\sin^22\,\theta}(2\,\xi_3^L-\frac{\sqrt{3}}{2}\sin^2\theta\,p_\phi)^2
+\frac{1}{4}\,p^2_\theta +\frac{1}{4}\,p^2_\phi\,,
\end{eqnarray}
or more explicitly in terms of the canonical coordinates
\begin{eqnarray}
 \mathrm{H}_{\scriptsize{\mathrm{SU(3)/SU(2)}}}
 &=&
\frac{1}{\sin^2{\theta}}
 \left( \frac{p_\alpha^2}{\sin^2\beta}
 +{p_\beta^2}+
 \left(\tan^2\,{\theta}+\frac{1}{\sin^2\beta}\right)p_\gamma^2
 -
2\,\frac{\cos\beta}{\sin^2\beta}\ p_\alpha p_\gamma +
\frac{\sqrt{3}}{2}\tan^2\theta\,{p_\gamma p_\phi} \right)
  \nn\\
   &+&\bigg.\frac{1}{4}\,p^2_\theta + \frac{1}{16}\left(
1+\frac{3}{\cos^2\,{\theta}}
   \right)\ p^2_\phi   \,.
\end{eqnarray}
Performing the inverse Legendre transformation we find the
Lagrangian
\begin{eqnarray}\label{eq:L32}
  \mathrm{L}_{\scriptsize{\mathrm{SU(3)/SU(2)}}} &=&\frac{1}{4}\,
\sin^2\,{\theta}
  \left(\big(1-\frac{1}{4}\cos^2\beta\,\sin^2\,{\theta}\big)
  \,\dot{\alpha}^2 +\dot{\beta}^2+
  \frac{1}{4}\left(3+\cos^2\,{\theta}\right)\,\dot{\gamma}^2\right.
  \nn\\
 &+&\bigg.\left.\frac{1}{2}\,\cos\beta\,\left(3+\cos^2\,{\theta}\right)\,
 \dot{\alpha}\dot{\gamma}-2{\sqrt{3}}\,(\cos\beta\,
\dot{\alpha} +\dot{\gamma})\dot\phi
  \right)+\dot\theta^2+\dot\phi^2\,.
\end{eqnarray}
Now one can consider the bilinear form  (\ref{eq:L32}) as the
metric $\mathbf{g}_{\mathcal{O}}$ on the orbit space
$\mathcal{O}=\mathrm{SU(3)/SU(2)}$
\begin{eqnarray}\label{eq:orbimetri}
  \mathbf{g}_{\mathcal{O}} &=&
  \frac{1}{4}\,
\sin^2\,{\theta}
  \left(\big(1-\frac{1}{4}\cos^2\beta\,\sin^2\,{\theta}\big)
  \,\mathrm{d}{\alpha}\otimes\mathrm{d}\alpha +\mathrm{d}{\beta}
\otimes\mathrm{d}\beta
  +
  \frac{1}{4}\left(3+\cos^2\,{\theta}\right)\mathrm{d}{\gamma}
  \otimes\mathrm{d}\gamma\right.
  \\
 &+&\bigg.\left.\frac{1}{2}\,\cos\beta\,
 \left(3+\cos^2\,{\theta}\right)\,
 \mathrm{d}{\alpha}\otimes\mathrm{d}{\gamma}
 -2{\sqrt{3}}\,(\cos\beta\,
\mathrm{d}{\alpha} +\mathrm{d}{\gamma})\otimes\mathrm{d}\phi
  \right)+\mathrm{d}\theta\otimes\mathrm{d}\theta+
  \mathrm{d}\phi\otimes\mathrm{d}\phi\,.\nn
\end{eqnarray}
Using our previous calculations (\ref{eq:volumeS5}) of
$\mathrm{Vol}(\mathbb{S}^5)$ with respect to the metric
(\ref{eq:S5eu}) induced by the canonical projection to the left
coset $\pi: \mathrm{SU(3)} \rightarrow
\mathrm{SU(3)}/\mathrm{SU(2)}$ and noting that the determinant of
the new  orbit metric (\ref{eq:orbimetri}) induced by the
Hamiltonian reduction  is
\begin{equation}\label{eq:detsu3su2}
\det \mathbf{g}_{\mathcal{O}}\,= \,\frac{1}{64}
\sin^{6}(\theta)\cos^2(\theta)\sin^{2}(\beta)\,,
\end{equation}
we find
\begin{equation}\label{eq:volrel}
    \mathrm{Vol}({\mathrm{SU(3)/SU(2)}})=\frac{\sqrt{3}}{2}\,
\mathrm{Vol}(\mathbb{S}^5)\,,
\end{equation}
with the same   \emph{stretching} factor $\sqrt{3}/{2}$ as found
in (\ref{eq:volumeSU3Mcd}) for the bi-invariant volume of the
SU(3) group.

\section{Riemannian structures on the quotient space  }

Now we are ready to answer the questions about the relation
between metric (\ref{eq:S5eu}) induced on the left coset
SU(3)/SU(2) by canonical projection from the ambient Euclidian
space and the metric (\ref{eq:orbimetri}) obtained as a result of
performing the Hamiltonian reduction  of the geodesic motion from
SU(3) to SU(3)/SU(2).

Performing a straightforward calculation of the Riemannian
curvature with respect to the metric (\ref{eq:orbimetri}) yields
\begin{equation}\label{eq:RsSu2/su3}
    \mathcal{R}\bigg(\mathbf{g}_{{}_{\scriptsize\mathrm{\frac{SU(3)}{SU(2)}}}}\bigg)
    = 21\,,
\end{equation}
while, from the embedding argumentation we used before,
 the Riemann scalar of the
unit five-sphere $\mathbb{S}^5$ with standard metric induced from
the Euclidean space is
\begin{equation}\label{eq:S5R}
    \mathcal{R}\bigg(\mathbf{g}_{{}_{\scriptsize\mathbb{S}^5}}\bigg)
    =20\,.
\end{equation}
Furthermore, even though  the Riemann scalar is a constant,
calculations shows that the metric (\ref{eq:L32}) {\it is not the
metric of a space of constant curvature}.

So, we have found that the Lagrangian of the reduced system defines
local flows on the configuration space which are not isometric to
those on $\mathbb{S}^5$ with its standard round metric.

We have shown above that the orbit space SU(3)/SU(2) considered as
a Riemannian space with metric  $\mathbf{g}$ induced from the
Cartan-Killing metric on SU(3) is not isometric to  the
$\mathbb{S}^5$ with the standard round metric
$\mathbf{g}_{\,\scriptsize\mathrm{S}^5}$. The next natural
question is  whether the metrics $\mathbf{g}$ and $\mathbf{g}_{\,
\scriptsize\mathrm{S}^5}$ are {\it geodesically }/{\it
projectively } equivalent.

There are several criteria on metrics to be geodesically
equivalent. According to L.P. Eisenhart \cite{Eisenhart},  two
metrics $\mathbf{g}$ and $\overline{\mathbf{g}}$ on
$n$-dimensional Riemann manifold {\it are geodesically equivalent
if and only if}
\begin{equation}\label{eq:geoequivcr}
2\,(n+1)\nabla_i(\mathbf{g})\,\overline{\mathbf{g}}_{jk}=
2\overline{\mathbf{g}}_{jk}\,\partial_i\Lambda +
\overline{\mathbf{g}}_{ik}\,\partial_j \Lambda +
\overline{\mathbf{g}}_{ji}\,\partial_k\Lambda \,,
\end{equation}
where $\nabla_i(\mathbf{g})$ is covariant with respect the metric
$\mathbf{g}$ an the scalar function $\Lambda$ is
\begin{equation}\label{eq:dd}
    \Lambda=\ln\left(
    \frac{\det(\overline{\mathbf{g}})}{\det(\mathbf{g})}
    \right)\,.
\end{equation}
According to our calculations
$$\det(\mathbf{g}_{\,\mathcal{O}})=
\frac{3}{4}\det(\mathbf{g}_{\,\mathbb{S}^5})$$ and
$$
\nabla_i\big(\mathbf{g}_{\,\mathbb{S}^5}\big)\
{\mathbf{g}_{\,\mathcal{O}}}_{\,jk}\neq 0\,,
$$
and therefore $\mathbf{g}_{\,\mathbb{S}^5}$ and
$\mathbf{g}_{\,\mathcal{O}}$  are not {\it geodesically }/{\it
projectively } equivalent.

\section{Conclusion}

In this paper we have presented, for the first time, the explicit
Hamiltonian  reduction from free motion on SU(3) to motion on the
coset space SU(3)/SU(2)$\approx\mathbb{S}^5$. This has been made
possible through a consistent parametrisation of SU(3) that
generalises the Euler angle parametrisation of SU(2). The full
details for this parametrisation of SU(3) are, for completeness,
collected together in an appendix to this paper. The results
presented there  have been  checked independently  using the
computer algebra packages  {\em Mathematica 5.0}  and {\em Maple
9.5}.

Through this analysis we have seen that the resulting dynamics is
not equivalent to the geodesic motion on $\mathbb{S}^5$ induced
from its standard round metric. This result prompts the following
questions.

\begin{itemize}
    \item Is it possible to identify, a priori, the induced metric
    on the coset space in terms of the properties of SU(3)?
    \item Is it possible to formulate the dynamics on SU(3) so that
    the reduced dynamics \emph{is} the expected geodesic motion on
    $\mathbb{S}^5$?
    \item What happens if we reduce to a non-zero level set of the
    integrals~(\ref{eq:SU20})?
\end{itemize}
Progress in answering these questions will, we feel,  throw  much
light on the dynamical aspects of the Hamiltonian reduction
procedure and hence lead to a deeper understanding of the
quantisation of  gauge theories.

\section*{Acknowledgments}

Helpful discussions during the work on the paper with T.~Heinzl,
D.~Mladenov  and  O.~Schr\"{o}der are acknowledged.

The contribution of  V.G., A.K. and Yu.P. was supported in part by
the Grant 04-01-00784 from the Russian Foundation for Basic
Research.

\newpage

\appendix
\section{Appendix}

\subsection{The su(3) algebra structure }
\label{sec:Append1}

The eight traceless $3\times 3$ Gell-Mann matrices providing a
basis for the su(3) algebra are listed below
\begin{equation}\label{lam-matr}
\begin{array}{c}
\begin{array}{ccc}
\lambda _{1}=\left(\begin{array}{ccc}
   0 & 1 & 0 \\
   1 & 0 & 0 \\
   0 & 0 & 0 \
 \end{array}\right),&
 \lambda _{2}=\left(\begin{array}{ccc}
   0 & -i & 0 \\
   i & 0 & 0 \\
   0 & 0 & 0 \
 \end{array}\right),&
  \lambda _{3}=\left(\begin{array}{ccc}
   1 & 0 & 0 \\
   0 & -1 & 0 \\
   0 & 0 & 0 \
 \end{array}\right),\\ & & \\
  \lambda _{4}=\left(\begin{array}{ccc}
   0 & 0 & 1 \\
   0 & 0 & 0 \\
   1 & 0 & 0 \
 \end{array}\right),&
 \lambda _{5}=\left(\begin{array}{ccc}
   0 & 0 & -i \\
   0 & 0 & 0 \\
   i & 0 & 0 \
 \end{array}\right),&
  \lambda _{6}=\left(\begin{array}{ccc}
   0 & 0 & 0 \\
   0 & 0 & 1 \\
   0 & 1 & 0 \
 \end{array}\right),\
 \end{array}
 \\ \\
\begin{array}{cc}
\lambda _{7}=\left(\begin{array}{ccc}
   0 & 0 & 0 \\
   0 & 0 & -i \\
   0 & i & 0 \
 \end{array}\right),&
  \lambda _{8}=\displaystyle{\frac{1}{\sqrt{3}}}\left(
  \begin{array}{ccc}
   1 & 0 & 0 \\
   0 & 1 & 0 \\
   0 & 0 & -2 \
\end{array}\right).
\end{array}\
\end{array}
\end{equation}
Sometimes it is convenient to use instead of the Gell-Mann
matrices the anti-Hermitian basis
$\mathbf{t}_a:=\dfrac{1}{2i}\lambda_a\,,$ obeying the relations
\begin{eqnarray}\label{eq:tt}
&& \mathbf{t}_a\,\mathbf{t}_b \,=\, - \frac{1}{6}\,\delta_{ab}\,
\mathbf{I}+
    \frac{1}{2}\,\sum_{c\,=\,1}^8\,\left(\, \mathrm{f}_{abc} -
    \imath\,\mathrm{d}_{abc}\right)\,
    \mathbf{t}_c\,,
\end{eqnarray}
where the structure constants $\mathrm{d}_{abc}$ are symmetric in
their indices  and the non-vanishing values are given in the Table
I, the coefficients $\mathrm{f}_{abc}$ are skew symmetric in all
indices. The constants $\mathrm{f}_{abc}$ determine the
commutators between the basis elements
\begin{equation}\label{eq:[tt]}
[\mathbf{t}_a\,,\mathbf{t}_b]\, =\,
   \sum_{c\, =\, 1}^{8}\, \mathrm{f}_{abc}\,\mathbf{t}_c\,.
\end{equation}
\begin{center}
\centerline{Table I. The symmetric coefficients
$\mathrm{d}_{abc}$}

\begin{tabular}{|c||c|c|c|c|}
  \hline
  $(abc)$
  &(118)(228)(338)
  &(146),(157)(256)(344)(355)
  &(247)(366)(377)&(448)(558)(668)(778)(888)
   \\
  \hline
   \hline
   &  &  &  &      \\
  $ \mathrm{d}_{abc} $ &$\dfrac{1}{\sqrt3} $&$\dfrac{1}{2} $ &$ -\dfrac{1}{2} $&
  $-\dfrac{1}{2\sqrt3}$ \\
 &  &  &  &   \\
  \hline
\end{tabular}
\end{center}
\begin{center}
\centerline{ Table II Structure of the su(3) algebra}
\bigskip
\bigskip
\begin{tabular}{|c||c|c|c|c|c|c|c|c|}
  \hline
   & $ \mathbf{t}_1$ & $ \mathbf{t}_2$ &$ \mathbf{t}_3$ &$ \mathbf{t}_4$ &$ \mathbf{t}_5$&
                                $\mathbf{t}_6$ & $\mathbf{t}_7 $& $\mathbf{t}_8$ \\
  \hline\hline
&&&&&&&  &  \\
 $\mathbf{t}_1$ & 0 & $ \mathbf{t}_3$ & $-\mathbf{t}_2$ & $\dfrac{1}{2}\,\mathbf{t}_7$ & $-\dfrac{1}{2}\,\mathbf{t}_6$
& $\dfrac{1}{2}\,\mathbf{t}_5$ & $-\dfrac{1}{2}\,\mathbf{t}_4$& 0 \\
&&&&&&&  &  \\
\hline
&&&&&&&  &  \\
 $\mathbf{t}_2$ &$-\mathbf{t}_3$ & 0 & $\mathbf{t}_1$ & $\dfrac{1}{2}\,\mathbf{t}_6$ & $\dfrac{1}{2}\,\mathbf{t}_7$
& $-\dfrac{1}{2}\,\mathbf{t}_4$ &  $-\dfrac{1}{2}\,\mathbf{t}_5$& 0  \\
&&&&&&&  &  \\
\hline
&&&&&&&  &  \\
$t _3 $& $ \mathbf{t}_2 $& $ -\mathbf{t}_1$&$0$&
$\dfrac{1}{2}\,\mathbf{t}_5$ & $-\dfrac{1}{2}\,\mathbf{t}_4$
& $-\dfrac{1}{2}\,\mathbf{t}_7$ & $\dfrac{1}{2}\,\mathbf{t}_6$& 0 \\
&&&&&&&  &  \\
 \hline
 &&&&&&&  &  \\
$\mathbf{t}_4$ & $-\dfrac{1}{2}\,\mathbf{t}_7 $&
$-\dfrac{1}{2}\,\mathbf{t}_6 $ & $-\dfrac{1}{2}\,\mathbf{t}_5 $
 & 0 &$ \frac{1}{2}\,\mathbf{t}_3 +\frac{\sqrt3}{2}\,\mathbf{t}_8$ & $\dfrac{1}{2}\,\mathbf{t}_2 $
& $\dfrac{1}{2}\,\mathbf{t}_1 $ &$-\dfrac{\sqrt3}{2}\,\mathbf{t}_5 $  \\
 &&&&&&&  &  \\
\hline
 &&&&&&&  &  \\
$\mathbf{t}_5$ & $\dfrac{1}{2}\,\mathbf{t}_6 $ &
$-\dfrac{1}{2}\,\mathbf{t}_7 $ & $\dfrac{1}{2}\,\mathbf{t}_4 $ &
$-\dfrac{1}{2}\,\mathbf{t}_3-\dfrac{\sqrt3}{2}\,\mathbf{t}_8$ & 0
& $-\dfrac{1}{2}\,\mathbf{t}_1 $
& $\dfrac{1}{2}\,\mathbf{t}_2 $ &  $\dfrac{\sqrt3}{2}\,\mathbf{t}_4 $\\
 &&&&&&&  &  \\
\hline
&&&&&&&  &  \\
 $\mathbf{t}_6$ &$-\dfrac{1}{2}\,\mathbf{t}_5 $  & $\dfrac{1}{2}\,\mathbf{t}_4 $
& $\dfrac{1}{2}\,\mathbf{t}_7 $ & $-\dfrac{1}{2}\,\mathbf{t}_2 $ &
$\dfrac{1}{2}\,\mathbf{t}_1 $ & 0 &
$-\dfrac{1}{2}\,\mathbf{t}_3+\dfrac{\sqrt3}{2}\,\mathbf{t}_8$
&  $-\frac{\sqrt3}{2}\,\mathbf{t}_7$  \\
&&&&&&&  &  \\
\hline
 &&&&&&&  &  \\
  $\mathbf{t}_7$ &  $\dfrac{1}{2}\,\mathbf{t}_4 $  & $\dfrac{1}{2}\,\mathbf{t}_5 $  & $-\dfrac{1}{2}\,\mathbf{t}_6 $
& $-\dfrac{1}{2}\,\mathbf{t}_1 $   & $-\dfrac{1}{2}\,\mathbf{t}_2
$&$\dfrac{1}{2}\,\mathbf{t}_3-\dfrac{\sqrt3}{2}\,\mathbf{t}_8$
& 0 & $\dfrac{\sqrt3}{2}\,\mathbf{t}_6 $  \\
 &&&&&&&  &  \\
\hline
 &&&&&&&  &  \\
 $\mathbf{t}_8$ & 0 & 0 & 0 & $ \dfrac{\sqrt3}{2}\,\mathbf{t}_5$ &
$-\dfrac{\sqrt3}{2}\,\mathbf{t}_4 $
&$\dfrac{\sqrt3}{2}\,\mathbf{t}_7 $&
$-\dfrac{\sqrt3}{2}\,\mathbf{t}_6$ &
0\\
 &&&&&&&  &  \\
\hline
\end{tabular}
\end{center}

\subsection{The basis of invariant 1-forms on the SU(3) group}

\label{sec:Append2}
\subsubsection{The left-invariant 1-forms}

Using the generalised Euler decomposition (\ref{eq:ESU(3)}) for
the SU(3) group element, it is straightforward to calculate the
left and right invariant 1-forms. The results  are given below
\noindent
\begin{itemize}
\item[]$
\begin{array}{l}\hspace*{-0.9cm}
{\omega}^1_L= \bigg(\cos[\beta]\sin[b]\cos[c]
(1-\displaystyle{\frac{1}{2}}\sin^2[\theta ])\\
\noalign{\vspace{0.90625ex}} \hspace{2.em} +\cos [\theta ]
\sin[\beta]\big(\cos [b]\cos[c]\cos[a+\gamma]-\sin[c]
\sin[a+\gamma]\big)\bigg){\mathrm{d}\alpha} \\
\noalign{\vspace{0.666667ex}} \hspace{2.em} -\cos [\theta
]\bigg(\cos[a+\gamma]\sin[c]+\cos[b]\cos[c]\sin [a+\gamma]\bigg){\mathrm{d}\beta} \\
\noalign{\vspace{0.916667ex}} \hspace{2.em} + \cos [c]\sin
[b]\big(1-\displaystyle{\frac{1}{2}}\sin^2[\theta
]\big){\mathrm{d}\gamma} + \cos [c] \sin [b] {\mathrm{d}a}-\sin[c]
{\mathrm{d}b}\,,
\end{array} $
\item[]$
\begin{array}{l}\hspace*{-0.9cm}
{\omega }^2_L= \bigg( \cos [\beta]\sin[b]\sin[c]
(1-\displaystyle{\frac{1}{2}}\sin^2[\theta ])  \\
\noalign{\vspace{0.90625ex}} \hspace{2.em} +\cos[\theta ] \sin [
\beta ]  \big(\cos [b] \cos [a+\gamma ] \sin [c]+\cos [c] \sin
[a+\gamma]\big)\bigg)
{\mathrm{d}\alpha}  \\
\noalign{\vspace{0.666667ex}} \hspace{2.em} +\cos [\theta ]
\bigg(\cos [c] \cos [a+\gamma]-\cos [b] \sin [
c] \sin [a+\gamma ]\bigg) {\mathrm{d}\beta} \\
\noalign{\vspace{0.916667ex}} \hspace{2.em} + \sin [b] \sin [
c](1-\displaystyle{\frac{1}{2}}\sin^2[\theta ])
{\mathrm{d}\gamma}+
  \sin [b] \sin [c] {\mathrm{d}a}+\cos[c]{\mathrm{d}b}\,,
\end{array}$
\item[]$
\begin{array}{l}\hspace*{-0.9cm}
{\omega }^3_L=\bigg(\cos[b] \cos [\beta ]
(1-\displaystyle{\frac{1}{2}}\sin^2[\theta ])-
 \cos [a+\gamma] \cos [\theta ] \sin [b] \sin [\beta ]\bigg) {\mathrm{d}\alpha} \\
\noalign{\vspace{0.916667ex}} \hspace{2.em}+ \cos [\theta ] \sin
[b] \sin [a+\gamma ] {\mathrm{d}\beta}+
 \cos [b] (1-\displaystyle{\frac{1}{2}}\sin^2[\theta ]){\mathrm{d}\gamma}
 +\cos[b]{\mathrm{d}a}+{\mathrm{d}c}\,,
\end{array}$
\item[]$
\begin{array}{l}\hspace*{-0.9cm}
{\omega }^4_L=\sin[\theta
]\bigg(\cos[\beta]\cos[\theta]\cos[\dfrac{b}{2}]
\cos\big[\dfrac{a+c}{2}+{\sqrt{3}}\phi\big]
-\cos \big[\dfrac{a-c}{2}+\gamma-{\sqrt{3}} \phi\big]\sin
[\dfrac{b}{2}]\sin [\beta]\bigg){\mathrm{d}\alpha}
\\
\noalign{\vspace{0.947917ex}} \hspace{2.em} + \sin [\dfrac{b}{2}]
\sin [\theta ] \sin \big[\dfrac{a-c}{2}+\gamma -{\sqrt{3}} \phi
\big] {\mathrm{d}\beta}
 \\
\noalign{\vspace{0.916667ex}} \hspace{2.em}+
\displaystyle{\frac{1}{2}}  \cos [\dfrac{b}{2}] \cos
\big[\dfrac{a+c}{2}+{\sqrt{3}} \phi \big] \sin [2 \theta ]
{\mathrm{d}\gamma}-2\,\cos [\dfrac{b}{2}] \sin
\big[\dfrac{a+c}{2}+{\sqrt{3}} \phi \big]{\mathrm{d}\theta}\,,
\end{array}$
\item[]$
\begin{array}{l}\hspace*{-0.9cm}
{\omega}^5_L=\sin[\theta ] \bigg(\sin [\dfrac{b}{2}]\sin[\beta ]
\sin\big[\dfrac{a-c}{2}+\gamma -{\sqrt{3}} \phi \big]
+\cos[\dfrac{b}{2}]\cos[\beta]\cos[\theta]\sin\big[\dfrac{a+c}{2}+
{\sqrt{3}}\phi \big]\bigg)
{\mathrm{d}\alpha}\\
\noalign{\vspace{0.708333ex}} \hspace{2.em} + \cos
\big[\dfrac{a-c}{2}+ \gamma -{\sqrt{3}} \phi \big]
\sin[\dfrac{b}{2}] \sin[\theta ]
{\mathrm{d}\beta} \\
\noalign{\vspace{0.916667ex}} \hspace{2.em}+\displaystyle{
\frac{1}{2}}  \cos [\dfrac{b}{2}] \sin [2 \theta ] \sin
\big[\dfrac{a+c}{2}+{\sqrt{3}} \phi \big]
{\mathrm{d}\gamma}+2\,\cos[\dfrac{b}{2}]\cos\big[\dfrac{a+c}{2}+
{\sqrt{3}}\phi \big]{\mathrm{d}\theta}\,,
\end{array}$
\item[]$
\begin{array}{l}\hspace*{-0.9cm}
{\omega }^6_L=\sin [\theta ] \bigg(\cos [\beta ] \cos [\theta ]
\cos \big[\dfrac{a-c}{2}+{\sqrt{3}} \phi \big] \sin[\dfrac{b}{2}]
+\sin[\beta]\cos\big[\dfrac{a+c}{2}+\gamma-{\sqrt{3}}\phi\big]
\cos[\dfrac{b}{2}]
\bigg)
{\mathrm{d}\alpha}\\
\noalign{\vspace{0.708333ex}} \hspace{2.em}- \cos [\dfrac{b}{2}]
\sin
[\theta ] \sin \big[\dfrac{a+c}{2}+\gamma -{\sqrt{3}} \phi \big] {\mathrm{d}\beta} \\
\noalign{\vspace{0.916667ex}} \hspace{2.em}+
\displaystyle{\frac{1}{2}}\cos \big[\dfrac{a-c}{2}+{\sqrt{3}} \phi
\big] \sin [\dfrac{b}{2}] \sin [2 \theta ]
{\mathrm{d}\gamma}-2\,\sin[\dfrac{b}{2}]\sin\big[\dfrac{a-c}{2}+{\sqrt{3}}
\phi\big]{\mathrm{d}\theta}\,,
\end{array}$
\item[]$
\begin{array}{l}\hspace*{-0.9cm}
{\omega}^7_L=\sin [\theta ]
\bigg(\cos[\beta]\cos[\theta]\sin[\dfrac{b}{2}] \sin
\big[\dfrac{a-c}{2}+{\sqrt{3}}\phi\big]-\cos [\dfrac{b}{2}] \sin[
\beta ] \sin \big[\dfrac{a+c}{2}+\gamma -{\sqrt{3}} \phi \big]
\bigg){\mathrm{d}
\alpha}\\
\noalign{\vspace{0.947917ex}} \hspace{2.em} -\cos[\dfrac{b}{2}]
\cos
\big[\dfrac{a+c}{2}+\gamma -{\sqrt{3}} \phi \big] \sin [\theta ]{\mathrm{d}\beta}\\
\noalign{\vspace{0.916667ex}} \hspace{2.em}+\displaystyle{
\frac{1}{2}}\sin [\dfrac{b}{2}] \sin [2 \theta ] \sin
\big[\dfrac{a-c}{2}+{\sqrt{3}} \phi \big] {\mathrm{d}\gamma}+
2\,\cos \big[\dfrac{a-c}{2}+{\sqrt{3}} \phi \big] \sin
[\dfrac{b}{2}] {\mathrm{d}\theta}\,,
\end{array}$
\item[]$ \hspace*{-0.9cm}{\omega}^8_L =-\displaystyle{ \frac{\sqrt{3}}{2}}  \cos
[\beta ] {{\sin^2 [\theta ]}} {\mathrm{d}\alpha}- \displaystyle{
\frac{\sqrt{3}}{2}}  {{\sin^2 [\theta ]}}
{\mathrm{d}\gamma}+2\,{\mathrm{d}\phi}\,. $
\end{itemize}

\subsubsection{The right-invariant 1-forms}

\begin{itemize}
\item[]$
\begin{array}{l}\hspace*{-0.9cm}
{\omega }^1_R =  \sin [\alpha ] {\mathrm{d}\beta}-\cos [\alpha ]
\sin [\beta ] {\mathrm{d}\gamma}-\cos [\alpha ] \sin [\beta ]
(1-\displaystyle{\frac{1}{2}}\sin^2[\theta ])\,{\mathrm{d}a}  \\
\noalign{\vspace{0.90625ex}} \hspace{2.em} + \cos [\theta ]
\bigg(\cos [a+\gamma]\sin [\alpha ]+\cos [\alpha ]\cos[\beta
] \sin [a+\gamma ]\bigg) {\mathrm{d}b} \\
\noalign{\vspace{0.916667ex}} \hspace{2.em} + \bigg(\cos[\theta ]
\sin [b]\big(-\cos [\alpha ] \cos [\beta ] \cos [a+\gamma] +
\sin [\alpha ]  \sin [a+\gamma] \big)\\
\noalign{\vspace{0.90625ex}} \hspace{2.em} - \cos [\alpha ]\cos[
b]  \sin [\beta] (1-\displaystyle{\frac{1}{2}}\sin^2[\theta
])\bigg){\mathrm{d}c} +{\sqrt{3}} \cos [\alpha] \sin [\beta
]{{\sin^2[\theta ]}} {\mathrm{d}\phi}\,,
\end{array}$
\medskip\item[]$
\begin{array}{l}\hspace*{-0.9cm}
{\omega }^2_R= \cos [\alpha ] {\mathrm{d}\beta}+\sin [\alpha ]
\sin [\beta ] {\mathrm{d}\gamma}+\sin [\alpha ] \sin [\beta ]
\big(1-\displaystyle{\frac{1}{2}}\sin^2[\theta ]\big) {\mathrm{d}a} \\
\noalign{\vspace{0.90625ex}} \hspace{2.em}+ \cos [\theta ]
\bigg(\cos [\alpha ] \cos [a+\gamma]-\cos [\beta ] \sin [\alpha
] \sin [a+\gamma ]\bigg) {\mathrm{d}b} \\
\noalign{\vspace{0.916667ex}} \hspace{2.em}+ \bigg( \cos [\theta ]
\sin [b] \big(\cos [\beta ] \cos [a+\gamma ] \sin [\alpha
]+\cos [\alpha ] \sin [a+\gamma ]\big)  \\
\noalign{\vspace{0.90625ex}} \hspace{2.em} +\cos [b]  \sin [
\alpha ] \sin [\beta ] (1-\displaystyle{\frac{1}{2}}\sin^2[\theta
])\bigg) {\mathrm{d}c}- \sqrt{3} \sin [\alpha ]
\sin [\beta ] {\sin^2[\theta ]}{\mathrm{d}\phi}\\
\end{array}$
\medskip\item[]$
\begin{array}{l}\hspace*{-0.9cm}
{\omega }^3_R={\mathrm{d}\alpha}+\cos [\beta ] {\mathrm{d}\gamma}+
\cos [\beta ] (1-\displaystyle{\frac{1}{2}}\sin^2[\theta ])
{\mathrm{d}a}+\cos
[\theta ] \sin [\beta] \sin [a+\gamma ] {\mathrm{d}b}  \\
\noalign{\vspace{1.15625ex}} \hspace{2.em}+ \bigg( \cos [b] \cos
[\beta ] (1-\displaystyle{\frac{1}{2}}\sin^2[\theta ])-\cos [
a+\gamma] \cos [\theta ] \sin [b] \sin [\beta ]\bigg)
{\mathrm{d}c}
-\sqrt{3}\cos[\beta]{{\sin^2[\theta ]}} {\mathrm{d}\phi}\,,
\end{array}$
\medskip\item[]$
\begin{array}{l}\hspace*{-0.9cm}
{\omega }^4_R= 2\,\cos [\dfrac{\beta}{2}]\sin [\dfrac{\alpha
+\gamma}{2} ] {\mathrm{d}\theta}- \displaystyle{\frac{1}{2}}
\cos[\dfrac{\beta}{2}] \cos [\dfrac{\alpha +\gamma}{2}] \sin [2
\theta ] {\mathrm{d}a}-\sin [\dfrac{\beta}{2} ] \sin
[a-\dfrac{\alpha -\gamma}{2} ] \sin [\theta ] {\mathrm{d}b} \\
\noalign{\vspace{1.15625ex}} \hspace{2.em}+
 \sin [\theta
]\bigg(\cos [a-\dfrac{\alpha -\gamma}{2} ] \sin [b] \sin
[\dfrac{\beta}{2} ] - \cos [b] \cos [\dfrac{\beta}{2} ]
\cos[\theta ] \cos [\dfrac{\alpha +\gamma}{2} ] \bigg)
{\mathrm{d}c}\\
\noalign{\vspace{1.15625ex}}\hspace{2.em} -\sqrt{3}\cos
[\dfrac{\beta}{2}] \cos [\dfrac{\alpha +\gamma}{2}]\sin [2\theta ]
{\mathrm{d}\phi}\,,
\end{array}$
\medskip\item[]$
\begin{array}{l}\hspace*{-0.9cm}
{\omega }^5_R=\cos [\dfrac{\beta}{2} ] \cos [\dfrac{\alpha
+\gamma}{2} ] {\mathrm{d}\theta}+ \displaystyle{\frac{1}{2}} \cos
[\dfrac{\beta}{2} ] \sin [\dfrac{\alpha +\gamma}{2} ] \sin [2
\theta ] {\mathrm{d}a}+\cos [a-\dfrac{\alpha
-\gamma}{2} ] \sin [\dfrac{\beta}{2} ] \sin [\theta ] {\mathrm{d}b} \\
\noalign{\vspace{1.15625ex}} \hspace{2.em}+ \sin [\theta ]
\bigg(\sin [b] \sin [\dfrac{\beta}{2} ] \sin [a-\dfrac{\alpha
-\gamma}{2} ]+ \cos [b] \cos [\dfrac{\beta}{2} ] \cos[\theta ]
\sin [\dfrac{\alpha +\gamma}{2} ]\bigg) {\mathrm{d}c}  \\
\noalign{\vspace{0.90625ex}} \hspace{2.em}+\sqrt{3}\cos
[\dfrac{\beta}{2} ] \sin [\dfrac{\alpha +\gamma}{2} ] \sin
[2\theta ]{\mathrm{d}\phi}\,,
\end{array}$
\medskip\item[]$
\begin{array}{l}\hspace*{-0.9cm}
{\omega }^6_R=2\,\sin [\dfrac{\beta}{2} ] \sin [\dfrac{\alpha
-\gamma}{2} ] {\mathrm{d}\theta}+ \displaystyle{\frac{1}{2}} \cos
[\dfrac{\alpha -\gamma}{2} ] \sin [\dfrac{\beta}{2} ] \sin [2
\theta ] {\mathrm{d}a}-\cos [\dfrac{\beta}{2} ] \sin
[a+\dfrac{\alpha +\gamma}{2} ] \sin [\theta ] {\mathrm{d}b}  \\
\noalign{\vspace{1.15625ex}} \hspace{2.em}+ \sin [\theta ]
\bigg(\cos [\dfrac{\beta}{2} ] \cos [ a+\dfrac{\alpha +\gamma}{2}
] \sin [b] + \cos [b] \cos [\theta ]\cos [\dfrac{\alpha -\gamma}{2} ]
\sin [\dfrac{\beta}{2} ] \bigg){\mathrm{d}c} \\
\noalign{\vspace{1.15625ex}} \hspace{2.em}+ \sqrt{3}\cos
[\dfrac{\alpha -\gamma}{2} ] \sin [\dfrac{\beta}{2} ] \sin
[2\theta ] {\mathrm{d}\phi}\,,
\end{array}$
\medskip\item[]$
\begin{array}{l}\hspace*{-0.9cm}
{\omega }^7_R=- 2\,\cos [\dfrac{\alpha -\gamma}{2} ] \sin
[\dfrac{\beta}{2} ] {\mathrm{d}\theta}+ \displaystyle{\frac{1}{2}}
\sin [\dfrac{\beta}{2}] \sin [\dfrac{\alpha -\gamma}{2} ] \sin [2
\theta ] {\mathrm{d}a}+ \cos [\dfrac{\beta}{2}]\cos
[a+\dfrac{\alpha +\gamma}{2} ] \sin [\theta ] {\mathrm{d}b} \\
\noalign{\vspace{1.15625ex}} \hspace{2.em}
 +  \sin [\theta]\bigg(\cos [\dfrac{\beta}{2}] \sin [b]\sin[a+
 \dfrac{\alpha +\gamma}{2} ]
 + \cos[b]\cos[\theta ]\sin[\dfrac{\beta}{2} ]
 \sin[\dfrac{\alpha -\gamma}{2} ]\bigg)
{\mathrm{d}c}\\
\noalign{\vspace{1.15625ex}}\hspace{2.em} +\sqrt{3}\sin
[\dfrac{\beta}{2}] \sin [\dfrac{\alpha -\gamma}{2} ] \sin [2
\theta ] {\mathrm{d}\phi}\,,
\end{array}$
\medskip\item[]$\hspace*{-0.9cm}
{\omega }^8_R=-\displaystyle{ \frac{\sqrt{3}}{2}} {{\sin^2[\theta
]}} {\mathrm{d}a}-\displaystyle{ \frac{\sqrt{3}}{2}} \cos [b]
{{\sin^2[\theta ]}} {\mathrm{d}c}+(2-{3}\sin^2[\theta ])
{\mathrm{d}\phi}\,.
 $
\end{itemize}

\subsection{The basis of the invariant vector fields on the SU(3) group }

The expressions for the left-invariant vector fields basis in the
Euler angles coordinate frame are given below

\subsubsection{The left-invariant vector fields}

\begin{itemize}
\item[]$
\begin{array}{l}\hspace*{-0.9cm}
{X }_1^L= \displaystyle{\frac{\cos [c]}{\sin [b]}}
{\dfrac{\partial}{\partial a}}=\sin [c] {\dfrac{\partial}{\partial
b}}-\cot [b] \cos [c]{\frac{\partial}{\partial  c}}\,,
\end{array}$
\medskip\item[]$
\begin{array}{l}\hspace*{-0.9cm}
{X }_2^L=\displaystyle{\frac{ \sin [c]}{ \sin [b]}}
{\dfrac{\partial}{\partial a}}+\cos[c] {\dfrac{\partial}{\partial
b}}- \cot [b] \sin [c] {\frac{\partial}{\partial  c}}\,,
\end{array}$
\medskip\item[]$
\begin{array}{l}\hspace*{-0.9cm}
{X }_3^L={\dfrac{\partial}{\partial  c}}\,,
\end{array}$
\medskip\item[]$
\begin{array}{l}\hspace*{-0.9cm}
{X }_4^L=-\displaystyle{\frac{\sin [\dfrac{b}{2}]}{\sin [\beta ]
\sin [\theta ]}\cos \big[\dfrac{a-c}{2}+\gamma -{\sqrt{3}} \phi
\big]{\dfrac{\partial}{\partial\alpha}} }
+\displaystyle{\frac{\sin [\dfrac{b}{2}]}{\sin [\theta ] } \sin
\big[\dfrac{a-c}{2}+
\gamma -{\sqrt{3}} \phi \big] {\dfrac{\partial}{\partial\beta}}}\\
\noalign{\vspace{0.916667ex}} \hspace{2.em}
+\left(\displaystyle{\frac{\sin[\dfrac{b}{2}]}{\sin [\theta ]}\cot
[\beta ]\cos \big[\dfrac{a-c}{2}+\gamma -{\sqrt{3}} \phi \big]\!+
\!\frac{2\cos[\dfrac{b}{2}]}{\sin [2\theta ]}\cos
\big[\dfrac{a+c}{2}+{\sqrt{3}} \phi
\big]}\right){\dfrac{\partial}{\partial\gamma}}\\
\noalign{\vspace{0.916667ex}} \hspace{2.em}-
\dfrac{1}{2}\cos
[\dfrac{b}{2}] \sin \big[\dfrac{a+c}{2}+{\sqrt{3}} \phi \big]
{\dfrac{\partial}{\partial\theta}}
 \displaystyle{-\frac{1}{2}
 \left(\frac{\cot[\theta
 ]}{\cos[\dfrac{b}{2}]}+\cos[\dfrac{b}{2}]\tan[\theta]\right)
\cos \big[\dfrac{a+c}{2}+{\sqrt{3}} \phi \big]
{\dfrac{\partial}{\partial a}} } \\
\noalign{\vspace{0.708333ex}} \hspace{2.em}+ \cot [\theta ] \sin
[\dfrac{b}{2}] \sin \big[\dfrac{a+c}{2}+{\sqrt{3}} \phi \big]
{\dfrac{\partial}{\partial b}}-\displaystyle{\frac{\cot [\theta
]}{2\cos [\frac{b}{2}]}} \cos \big[\dfrac{a+c}{2}+{\sqrt{3}} \phi
\big]{\frac{\partial}{\partial  c}}
 \\
\noalign{\vspace{1.15625ex}} \hspace{2.em}
\displaystyle{+\frac{\sqrt{3}}{4} } \cos [\dfrac{b}{2}] \cos
\big[\dfrac{a+c}{2}+{\sqrt{3}} \phi \big]\tan [\theta ]
{\dfrac{\partial}{\partial\phi}}\,,
\end{array}$
\medskip\item[]$
\begin{array}{l}\hspace*{-0.9cm}
{X }_5^L=\displaystyle{\frac{\sin [\dfrac{b}{2}]}{\sin [\beta
]\sin [\theta ]}}\sin \big[\dfrac{a-c}{2}+\gamma -{\sqrt{3}} \phi
\big] {\dfrac{\partial}{\partial\alpha}} +
\displaystyle{\frac{\sin [\dfrac{b}{2}]}{\sin [\theta ]}} \cos
\big[\dfrac{a-c}{2}+
\gamma -{\sqrt{3}} \phi \big] {\dfrac{\partial}{\partial\beta}}\\
\noalign{\vspace{0.916667ex}} \hspace{2.em}
=\left(\displaystyle{\frac{\sin[\dfrac{b}{2}]}{\sin [\theta ]}\cot
[\beta ]\sin \big[\frac{a-c}{2}+\gamma -{\sqrt{3}} \phi \big]-
\frac{2\cos[\frac{b}{2}]}{\sin [2\theta ]}\sin
\big[\dfrac{a+c}{2}+{\sqrt{3}} \phi
\big]}\right){\dfrac{\partial}{\partial\gamma}}\\
\noalign{\vspace{1.15625ex}}
\hspace{2.em}+\dfrac{1}{2}\cos[\dfrac{b}{2}] \cos
\big[\dfrac{a+c}{2}+{\sqrt{3}} \phi \big]
{\dfrac{\partial}{\partial\theta}} \displaystyle{-\frac{1}{2}
\left(\frac{\cot[\theta
 ]}{\cos[\dfrac{b}{2}]}+\cos[\dfrac{b}{2}]\tan[\theta]\right)
\sin \big[\dfrac{a+c}{2}+{\sqrt{3}} \phi \big] {\dfrac{\partial}{\partial a}} } \\
\noalign{\vspace{0.916667ex}} \hspace{2.em}-\cos
\big[\dfrac{a+c}{2}+{\sqrt{3}} \phi \big] \cot [\theta ] \sin
[\dfrac{b}{2}] {\dfrac{\partial}{\partial b}}\displaystyle{
-\frac{\cot [\theta ]}{2\cos [\dfrac{b}{2}]}} \sin
\big[\dfrac{a+c}{2}+{\sqrt{3}} \phi \big]
{\frac{\partial}{\partial c}}
 \\
\noalign{\vspace{1.15625ex}} \hspace{2.em}
+\displaystyle{\frac{\sqrt{3}}{4}} \cos [\dfrac{b}{2}] \sin
\big[\dfrac{a+c}{2}+{\sqrt{3}} \phi \big] \tan [\theta
]{\dfrac{\partial}{\partial\phi}}\,,
\end{array}$
\medskip\item[]$
\begin{array}{l}\hspace*{-0.9cm}
{X }_6^L=\displaystyle{\frac{\cos [\dfrac{b}{2}]}{\sin [\beta ]
\sin [\theta ]}}\cos \big[\dfrac{a+c}{2}+\gamma -{\sqrt{3}} \phi
\big]{\dfrac{\partial}{\partial\alpha}}-\displaystyle{\frac{\cos
[\dfrac{b}{2}]}{\sin [\theta ]}} \sin \big[\dfrac{a+c}{2}+\gamma
-{\sqrt{3}} \phi \big]
{\dfrac{\partial}{\partial\beta}}\\
\noalign{\vspace{0.708333ex}}\hspace{2.em}
-\left(\displaystyle{\frac{\cos[\dfrac{b}{2}]}{\sin [\theta ]}\cot
[\beta ]\cos \big[\dfrac{a+c}{2}+\gamma -{\sqrt{3}} \phi \big]-
\frac{2\sin[\dfrac{b}{2}]}{\sin [2\theta ]}\cos
\big[\dfrac{a-c}{2}+{\sqrt{3}} \phi
\big]}\right) {\dfrac{\partial}{\partial\gamma}}\\
\noalign{\vspace{0.708333ex}}\hspace{2.em} - \dfrac{1}{2} \sin
[\dfrac{b}{2}] \sin \big[\dfrac{a-c}{2}+{\sqrt{3}} \phi \big]
{\dfrac{\partial}{\partial\theta}}\displaystyle{-\frac{1}{2}
\left(\frac{\cot[\theta
 ]}{\sin[\dfrac{b}{2}]}+\sin[\dfrac{b}{2}]\tan[\theta]\right)
\cos \big[\dfrac{a-c}{2}+{\sqrt{3}} \phi \big]
{\dfrac{\partial}{\partial a}} } \\
\noalign{\vspace{0.947917ex}} \hspace{2.em} -\cos [\dfrac{b}{2}]
\cot [\theta ] \sin \big[\dfrac{a-c}{2}+{\sqrt{3}} \phi \big]
{\dfrac{\partial}{\partial b}}+ \displaystyle{\frac{\cot [\theta
]}{2\sin [\dfrac{b}{2}]}\cos
\big[\dfrac{a-c}{2}+{\sqrt{3}} \phi \big]}
{\frac{\partial}{\partial  c}} \\
\noalign{\vspace{1.15625ex}} \hspace{2.em}+
\displaystyle{\frac{\sqrt{3}}{4} }
 \cos \big[\dfrac{a-c}{2}+{\sqrt{3}} \phi
\big] \sin [\dfrac{b}{2}] \tan [\theta
]{\dfrac{\partial}{\partial\phi}}\,,
\end{array}$
\medskip\item[]$
\begin{array}{l}\hspace*{-0.9cm}
{X }_7^L=-\displaystyle{\frac{\cos [\dfrac{b}{2}]}{\sin [\beta ]
\sin [\theta ]}}\sin \big[\dfrac{a+c}{2}+\gamma -{\sqrt{3}} \phi
\big] {\dfrac{\partial}{\partial\alpha}}
 -\displaystyle{\frac{\cos [\dfrac{b}{2}]}{ \sin [\theta ]}} \cos
\big[\dfrac{a+c}{2}+\gamma -{\sqrt{3}} \phi \big]
{\dfrac{\partial}{\partial\beta}}\\
 \noalign{\vspace{1.15625ex}} \hspace{2.em}
 +\left(\displaystyle{\frac{\cos[\dfrac{b}{2}]}{\sin [\theta ]}\cot
[\beta ]\sin \big[\dfrac{a+c}{2}+\gamma -{\sqrt{3}} \phi \big]+
\frac{2\sin[\dfrac{b}{2}]}{\sin [2\theta ]}\sin
\big[\dfrac{a-c}{2}+{\sqrt{3}} \phi
\big]}\right)  {\dfrac{\partial}{\partial\gamma}} \\
\noalign{\vspace{0.708333ex}} \hspace{2.em} + \dfrac{1}{2}\cos
\big[\dfrac{a-c}{2}+{\sqrt{3}} \phi \big] \sin [\dfrac{b}{2}]
{\dfrac{\partial}{\partial\theta}} \displaystyle{-\frac{1}{2}
\left(\frac{\cot[\theta
 ]}{\sin[\dfrac{b}{2}]}+\sin[\dfrac{b}{2}]\tan[\theta]\right)
\sin \big[\dfrac{a-c}{2}+{\sqrt{3}} \phi \big]}
{\dfrac{\partial}{\partial a}} \\
\noalign{\vspace{0.947917ex}} \hspace{2.em}+ \cos [\dfrac{b}{2}]
\cos \big[\dfrac{a-c}{2}+{\sqrt{3}} \phi \big] \cot [\theta ]
{\dfrac{\partial}{\partial b}}+\displaystyle{\frac{\cot [\theta
]}{2\sin [\dfrac{b}{2}]}}\sin
\big[\dfrac{a-c}{2}+{\sqrt{3}} \phi \big] {\frac{\partial}{\partial  c}} \\
\noalign{\vspace{1.15625ex}} \hspace{2.em}
+\displaystyle{\frac{\sqrt{3}}{4}}\sin [b] \sin
\big[\dfrac{a-c}{2}+{\sqrt{3}} \phi \big] \tan [\theta
]{\dfrac{\partial}{\partial\phi}}\,,
\end{array}$
\medskip\item[]$
\begin{array}{l}\hspace*{-0.9cm}
{X }_8^L= \dfrac{1}{2}\,{\dfrac{\partial}{\partial\phi}}\,.
\end{array}$
\end{itemize}

\subsubsection{The right-invariant  vector fields}

\begin{itemize}
\item[]$
\begin{array}{l}\hspace*{-0.9cm}
{X }_1^R=\cos [\alpha ] \cot [\beta ]
{\dfrac{\partial}{\partial\alpha}} +\sin [\alpha ]
{\dfrac{\partial}{\partial\beta}}-\displaystyle{\frac{\cos [
\alpha ]}{\sin [\beta ]}} {\dfrac{\partial}{\partial\gamma}}\,,
\end{array}$
\medskip\item[]$
\begin{array}{l}\hspace*{-0.9cm}
{X }_2^R=-\sin [\alpha ] \cot [\beta ]
{\dfrac{\partial}{\partial\alpha}}+\cos [\alpha ]
{\dfrac{\partial}{\partial\beta}}+\displaystyle{\frac{\sin [\alpha
]}{ \sin [\beta ]}} {\dfrac{\partial}{\partial\gamma}}\,,
\end{array}$
\medskip\item[]$
\begin{array}{l}\hspace*{-0.9cm}
{X }_3^R={\dfrac{\partial}{\partial\alpha}}\,,
\end{array}$
\medskip\item[]$
\begin{array}{l}\hspace*{-0.9cm}
{X }_4^R=\displaystyle{\frac{\cot [\theta ]}{2 \cos
[\dfrac{\beta}{2}]}} \cos [\dfrac{\alpha +\gamma}{2} ]
{\dfrac{\partial}{\partial\alpha}}-
 \cot [\theta ] \sin
[\dfrac{\beta}{2} ] \sin [\dfrac{\alpha +\gamma}{2}]
{\dfrac{\partial}{\partial\beta}}\\
\noalign{\vspace{0.916667ex}} \hspace{2.em} +\cos [\dfrac{\alpha
+\gamma}{2} ]\left(\displaystyle{\frac{\cot [\theta
]}{2\cos[\dfrac{\beta}{2}]}} -\cos [\dfrac{\beta}{2} ] \tan
[\theta ]\right) {\dfrac{\partial}{\partial\gamma}} + \dfrac{1}{2}
\cos [\dfrac{\beta}{2} ] \sin
[\dfrac{\alpha +\gamma}{2}] {\dfrac{\partial}{\partial\theta}}\\
\noalign{\vspace{0.916667ex}} \hspace{2.em} -\left(\displaystyle{
\frac{\cot [b]}{\sin [\theta ]}
 \cos [a-\dfrac{\alpha -\gamma}{2} ]\sin [\dfrac{\beta}{2} ]
 +\frac{\cos [\dfrac{\beta}{2} ]}{\sin
[2\theta ]}\cos [\dfrac{\alpha +\gamma}{2} ]
(2-3 \sin^2[\theta ])}\right) {\dfrac{\partial}{\partial a}}  \\
\noalign{\vspace{0.916667ex}} \hspace{2.em}-\,
\displaystyle{\frac{\sin [\dfrac{\beta}{2}]}{\sin [\theta ]}}
 \sin [a-\dfrac{\alpha -\gamma}{2} ] {\dfrac{\partial}{\partial b}}
 +\displaystyle{
 \frac{\sin [\dfrac{\beta}{2} ]}{ \sin [b] \sin [\theta ]}}
 \cos [ a-\dfrac{\alpha-\gamma}{2} ]  {\frac{\partial}{\partial  c}} \\
\noalign{\vspace{1.15625ex}}
\hspace{2.em}-\displaystyle{\frac{\sqrt{3}}{4}}  \cos
[\dfrac{\beta}{2} ] \cos [\dfrac{\alpha +\gamma}{2} ] \tan [\theta
]{\dfrac{\partial}{\partial\phi}}\,,
\end{array}$
\medskip\item[]$
\begin{array}{l}\hspace*{-0.9cm}
{X }_5^R=-\displaystyle{\frac{\cot [\theta ]}{2\cos
[\dfrac{\beta}{2} ]}} \sin [\dfrac{\alpha +\gamma}{2} ]
{\dfrac{\partial}{\partial\alpha}}-\cos [\dfrac{\alpha +\gamma}{2}
] \cot [\theta ] \sin [\dfrac{\beta}{2} ]
{\dfrac{\partial}{\partial\beta}}  \\
\noalign{\vspace{1.15625ex}} \hspace{2.em}- \sin [\dfrac{\alpha
+\gamma}{2} ]\left(\displaystyle{\frac{\cot [\theta
]}{2\cos[\dfrac{\beta}{2}]}} -\cos [\dfrac{\beta}{2} ]\tan [\theta
]\right) {\dfrac{\partial}{\partial\gamma}}+ \dfrac{1}{2}\,\cos
[\dfrac{\beta}{2} ] \cos [\dfrac{\alpha +\gamma}{2} ]
{\dfrac{\partial}{\partial\theta}}\\
\noalign{\vspace{0.916667ex}} \hspace{2.em}-\left(
\displaystyle{\frac{\cot [b]}{\sin [\theta ]}
 \sin [a-\dfrac{\alpha -\gamma}{2} ]\sin [\dfrac{\beta}{2}]-
 \frac{\cos [\dfrac{\beta}{2} ]}{\sin
[2\theta ]}\sin [\dfrac{\alpha +\gamma}{2} ] (2-3 \sin^2[\theta])}
\right){\dfrac{\partial}{\partial a}}\\
\noalign{\vspace{0.916667ex}} \hspace{2.em}+\, \displaystyle{
\frac{\sin [\dfrac{\beta}{2} ]}{\sin [\theta ]}\cos
[a-\dfrac{\alpha +\gamma}{2} ] } {\dfrac{\partial}{\partial b}}+
\displaystyle{\frac{ \sin [\dfrac{\beta}{2} ]}{\sin [b]\sin
[\theta ]} \sin [a-\dfrac{\alpha -\gamma}{2} ]}
{\frac{\partial}{\partial  c}} \\
\noalign{\vspace{1.15625ex}} \hspace{2.em}+
\displaystyle{\frac{\sqrt{3}}{4} } \cos [\dfrac{\beta}{2} ] \sin
[\dfrac{\alpha +\gamma}{2} ] \tan [\theta
]{\dfrac{\partial}{\partial\phi}}\,,
\end{array}$
\medskip\item[]$
\begin{array}{l}\hspace*{-0.9cm}
{X }_6^R=\displaystyle{\frac{\cot [\theta ]}{2\sin
[\dfrac{\beta}{2} ]}} \cos [\dfrac{\alpha -\gamma}{2} ]
{\dfrac{\partial}{\partial\alpha}} +\cos [\dfrac{\beta}{2} ]\cot
[\theta ]
\sin [\dfrac{\alpha-\gamma}{2}]  {\dfrac{\partial}{\partial\beta}}  \\
\noalign{\vspace{1.15625ex}} \hspace{2.em}-\cos [\dfrac{\alpha
-\gamma}{2}] \left(\displaystyle{\frac{\cot [\theta
]}{2\sin[\dfrac{\beta}{2}]}} -\sin [\dfrac{\beta}{2} ] \tan
[\theta ]\right) {\dfrac{\partial}{\partial\gamma}}+
\dfrac{1}{2}\, \sin [\dfrac{\beta}{2} ] \sin [\dfrac{\alpha
-\gamma}{2} ]
{\dfrac{\partial}{\partial\theta}}\\
\noalign{\vspace{0.916667ex}} \hspace{2.em}- \left(
\displaystyle{\frac{\cot [b]}{\sin [\theta ]}
 \cos [a+\dfrac{\alpha +\gamma}{2} ]\cos [\dfrac{\beta}{2} ]
 -\frac{\sin [\dfrac{\beta}{2} ]}{\sin
[2\theta ]}\cos [\dfrac{\alpha -\gamma}{2} ] (2-3 \sin^2[\theta
])}
\right){\dfrac{\partial}{\partial a}}\\
\noalign{\vspace{0.916667ex}} \hspace{2.em}-\, \displaystyle{
\frac{\cos [\dfrac{\beta}{2} ]}{\sin [\theta ]}\sin [a+
\dfrac{\alpha +\gamma}{2} ] } {\dfrac{\partial}{\partial b}} +
\displaystyle{\frac{ \cos [\dfrac{\beta}{2} ]}{\sin [b]\sin
[\theta ]} \cos [a+\dfrac{\alpha +\gamma}{2} ]}
{\frac{\partial}{\partial  c}} \\
\noalign{\vspace{1.15625ex}} \hspace{2.em}+
\displaystyle{\frac{\sqrt{3}}{4} }  \cos [\dfrac{\alpha
-\gamma}{2} ] \sin [\dfrac{\beta}{2} ]\tan [\theta
]{\dfrac{\partial}{\partial\phi}}\,,
\end{array}$
\medskip\item[]$
\begin{array}{l}\hspace*{-0.9cm}
{X }_7^R=\displaystyle{\frac{\cot [\theta ]}{2\sin
[\dfrac{\beta}{2} ]}} \sin [\dfrac{\alpha -\gamma}{2} ]
{\dfrac{\partial}{\partial\alpha}}-\cos [\dfrac{\beta}{2} ]\cos
[\dfrac{\alpha-\gamma}{2} ] \cot [\theta ]
{\dfrac{\partial}{\partial\beta}}  \\
\noalign{\vspace{1.15625ex}} \hspace{2.em}-\sin [\dfrac{\alpha
-\gamma}{2}]
 \left(\displaystyle{\frac{\cot [\theta ]}{2\sin[\dfrac{\beta}{2}]}}
 -\sin[\dfrac{\beta}{2} ]\tan [\theta ]\right)
 {\dfrac{\partial}{\partial\gamma}}-
 \dfrac{1}{2}\,
 \cos [\dfrac{\alpha -\gamma}{2} ]\sin [\dfrac{\beta}{2} ]
  {\dfrac{\partial}{\partial\theta}}\\
\noalign{\vspace{0.916667ex}} \hspace{2.em}-\left(
\displaystyle{\frac{\cot [b]}{\sin [\theta ]} \cos
[\dfrac{\beta}{2} ] \sin [a+\dfrac{\alpha +\gamma}{2} ]-
\frac{\sin [\dfrac{\beta}{2} ]}{\sin [2\theta ]}\sin
[\dfrac{\alpha -\gamma}{2} ] (2-3 \sin^2[\theta ])}
\right){\dfrac{\partial}{\partial a}}\\
\noalign{\vspace{0.916667ex}} \hspace{2.em}+\, \displaystyle{
\frac{\cos [\dfrac{\beta}{2} ]}{\sin [\theta ]}\cos
[a+\dfrac{\alpha +\gamma}{2} ] } {\dfrac{\partial}{\partial
b}}+\displaystyle{\frac{ \cos [\dfrac{\beta}{2} ]}{\sin [b]\sin
[\theta ]} \sin [a+\dfrac{\alpha +\gamma}{2} ]}
{\frac{\partial}{\partial  c}} \\
\noalign{\vspace{1.15625ex}} \hspace{2.em}+
\displaystyle{\frac{\sqrt{3}}{4} } \sin [\dfrac{\beta}{2} ]\sin
[\dfrac{\alpha -\gamma}{2} ]\tan [\theta
]{\dfrac{\partial}{\partial\phi}}\,,
\end{array}$
\medskip\item[]$
\begin{array}{l}\hspace*{-0.9cm}
{X }_8^R={\sqrt{3}} {\dfrac{\partial}{\partial\gamma}}-{\sqrt{3}}
{\dfrac{\partial}{\partial a}}+
\dfrac{1}{2}\,{\dfrac{\partial}{\partial\phi}}\,.
\end{array}$
\end{itemize}

\newpage


\end{document}